\documentclass[%
reprint,
superscriptaddress,
nofootinbib,
amsmath,amssymb,
aps,
prc,
floatfix,
showkeys,
]{revtex4-2}

\usepackage{graphicx}
\usepackage{dcolumn}
\usepackage{bm}
\usepackage{xcolor}
\usepackage[colorlinks=true,citecolor=blue,linkcolor=blue,anchorcolor=blue,filecolor=blue,
urlcolor=blue]{hyperref}
\usepackage{soul}

\begin{document}

\title{\textbf{Classification of $g$-modes for neutron stars with a strong transition: \\Novel universal relation including slow stable hybrid stars}} 

\author{M. C. Rodriguez}
\affiliation{Grupo de Astrofísica de Remanentes Compactos, Facultad de Ciencias Astronómicas y Geofísicas, Universidad Nacional de La Plata, Paseo del Bosque S/N, 1900, La Plata, Argentina.}
\affiliation{CONICET, Godoy Cruz 2290, 1425, CABA, Argentina.}

\author{José C. Jiménez}
\affiliation{Department of Astrophysics, The Brazilian Center for Research in Physics (CBPF), Rua Dr. Xavier Sigaud, 150, URCA, Rio de Janeiro CEP 22210-180, RJ, Brazil}
\affiliation{Universidad Tecnológica del Perú}

\author{Ignacio F. Ranea-Sandoval}
\affiliation{Grupo de Astrofísica de Remanentes Compactos, Facultad de Ciencias Astronómicas y Geofísicas, Universidad Nacional de La Plata, Paseo del Bosque S/N, 1900, La Plata, Argentina.}
\affiliation{CONICET, Godoy Cruz 2290, 1425, CABA, Argentina.}

\date{\today}

\begin{abstract}
We investigate the behavior of the non-radial gravity-pulsation discontinuity $g$-mode in neutron stars with a strong first-order phase transition which give rise to hybrid-star configurations. These modes are of utmost relevance since they can be potentially excited in isolated as well as binary neutron star systems in the inspiral phase, thus allowing us to indirectly detect the presence of a deconfinement transition. In order to do this, we consider four categories of hybrid stars that present distinctive features in their equations of state. We employ the constant speed of sound parametrization, which accounts for a sharp phase transition between confined hadronic matter and deconfined quark matter. Then, working within the relativistic Cowling approximation to obtain the frequencies of non-radial oscillations, we find that, depending on the hybrid star category, the relations between discontinuity $g$-mode frequencies and masses as well as tidal deformabilities display a highly distinct behavior across the diverse hybrid star categories that appear in the slow hadron-quark conversion regime. This distinct phenomenology provides smoking-gun evidence to clearly distinguish and further classify hybrid stars with a strong transition from purely hadronic stars using upcoming gravitational-wave data. In addition, we present for each of the categories studied the relation between the $g$-mode frequency and the normalized energy density jump. Finally, we present a novel universal relation for the discontinuity $g$-mode able to encompass the four categories including long branches of slow stable hybrid stars and address its asteroseismological capability.
\end{abstract}

\maketitle

\section{Introduction} \label{sec:intro}

Neutron stars (NSs) are compact objects formed in explosive environments, such as core-collapse supernovae \cite{glendenning2012compact}. Simulations show \cite{Muller:2016izw} that these dense objects are born with extremely fast rotation, large magnetic fields, and somewhat high temperatures. However, even if the strength of these features decrease as they get older\footnote{For supernovae, protoneutron stars cool down from $\sim$ 50 to 0.01 MeV within $10^2$ seconds through neutrino cooling \cite{Lattimer:2004tpo}.}, NSs can still be considered the most extreme systems in nature (aside from black holes), thus being also useful to probe dense nuclear matter \cite{Schaffner-Bielich:2020psc}. 

Except in fully dynamical situations like those occurring in the final stages of NS mergers \cite{Shankar:2025pwq}, one can safely assume that NSs are cold and spherically symmetric in most astrophysical scenarios aimed to study their morphological and oscillatory properties. In nature, they are mainly observed as pulsars with some of them having high masses ($ \sim 2~M_{\odot}$) and small radii ($\sim 12$~km) \cite{Fan:2023spm}. Both structural properties imply baryon densities, $n_B$, several times larger than the nuclear saturation one, \mbox{$n_0\sim 0.16 \, \rm{fm}^{-3}$}, at the core of NSs. 



In turn, this motivated the astrophysics community to hypothesize that at the inner core of NSs, such high densities might give rise to complex phases of nuclear matter, e.g. deconfined quark matter, only somewhat known in phenomenological models \cite{potekhin:2010tpo,baym:2018fht,orsaria:2019pti}. Unfortunately, current terrestrial laboratories are far from being able to simultaneously reach these high-density and low-temperature conditions. For this reason, NSs are natural astrophysical laboratories useful to understand the nature of matter in such extreme regimes of pressure and density \cite{baym:2018fht,orsaria:2019pti}.

Schematically, NSs can be thought of as being composed of three layers of nuclear matter with radically different behaviors: an inner core, an outer core, and a crust. Baryon densities in the crust are lower than $n_0$, so experimental data from laboratories can be used to reduce the uncertainties on the low-density regime of the NS matter equation of state (EoS). Moreover, calculations using Chiral Effective Theory (CET) have proven to be useful to constrain the EoS in such a density regime (see, e.g., Ref.~\cite{Drischler:2020hwd} and references therein). This situation changes in the outer core and, despite the help provided by new observational data from mergers of NSs, there is still no general agreement on the behavior of matter for densities above $n_0$. This lack of knowledge becomes more drastic as density increases at the inner core of such compact objects. In order to describe matter in the depths of the inner core, several theoretical lines are being explored. One of these alternatives includes the existence of a hadron-quark phase transition (see, e.g., Refs.~\cite{weber:2005sqm,baym:2018fht,orsaria:2019pti} and references therein), which might form the so-called hybrid star, meaning that they are composed of hadrons in the outer region while quarks dominate the stellar core.

On the other hand, observations of the $2~M_{\odot}$ pulsars PSR J1614-2230 \cite{Demorest:2010ats} (later improved in Ref. \cite{Arzoumanian:2018tny}), PSR J0348+0432 \cite{Antoniadis:2013amp} and PSR J0740+6620 \cite{Cromartie:2020rsd} impose strong constraints on the NS matter EoS. Other constraints become available from gravitational-wave events, particularly useful was GW170817 and its electromagnetic counterpart \citep{Abbott:2017gwa, Abbott:2018gmo, Abbott:2020goo}, which offered valuable insights into the EoS and the internal structure of NSs by constraining the dimensionless tidal deformability, $\Lambda$, of these compact objects. In addition, observations from NASA's NICER X-ray telescope \citep{Miller:2019pjm, Riley:2019anv, Miller:2021tro, Riley:2021anv,salmi:2024tro,choudhury:2024anv} have been used to determine the masses and radii of isolated NSs. 

In principle, the calculation of the NS EoS should come from employing quantum chromodynamics (QCD), the theory \textcolor{blue}{of} strong interactions. The main key features of QCD are confinement, chiral symmetry breaking and asymptotic freedom, which combined suggest that strongly interacting matter must undergo a phase transition in which hadrons liberate their internal components, i.e. quarks and gluons forming dense/hot quark matter.  {The nature and position in the QCD phase diagram of such hadron-quark phase transition remain one of the most important  {and challenging} open questions related to dense matter physics. Despite large uncertainties, many theoretical works agree that for cold matter, such a phase transition might occur at densities between $n_B \sim (5-10)\, n_0$ (see, for example, Refs. \cite{Lattimer:2004tpo,baym:2018fht,orsaria:2019pti,lattimer:2021nsa} and references therein).} In particular, performing lattice QCD calculations at finite baryon densities is extremely difficult (perhaps impossible) as the theory has intrinsic computational problems coming from the so-called ``fermionic sign problem'' when computing thermodynamic observables using quantum Monte Carlo techniques \citep{PhysRevLett.94.170201,PAN2024879}. This is one of the main reasons that led to the development of several phenomenological and effective models that could reproduce and yield a proper understanding of (at least some) key features of QCD while keeping (or breaking as  {happens} with the chiral symmetry) some of its underlying symmetries (see, e.g., Refs.~\cite{baym:2018fht,orsaria:2019pti} and references therein).

In this sense, one of the most challenging open questions in extreme QCD physics is connected to determining the existence and nature of a potential phase transition to deconfined quark matter at intermediate baryon densities and its possible realization at the inner core of (high-mass) NSs. Interestingly, this complex phenomenon could also leave unique imprints not only in {\it quasi-static} observables (e.g. masses, radii, tidal deformabilities), but also in {\it oscillatory} ones, such as their non-radial oscillation modes
. They might be excited, e.g., by tidal forces in NS binaries during the late stages of the inspiral phase as the orbital frequency approaches the star's normal oscillation frequencies. During this stage, different non-radial modes, such as the composition gradient $g$-modes, can be resonantly excited \citep{10.1093/mnras/222.3.393,10.1093/mnras/227.2.265}. These non-radial oscillations are then damped out by the emission of gravitational waves carrying information about the stellar interiors. In particular, it has been argued somewhat recently in Ref. \cite{counsell:2025imi} that discontinuity $g$-modes (similar to the ones explored in the present work) might present tidal coupling, thus being potentially observable in the gravitational waveforms during the inspiral stage (see discussion later in this section). This opens a window with the prospective to characterize the unknown microphysics (bound to phase transitions). 


Currently, one intriguing possibility for this hypothesized hadron-quark phase transition at low temperatures and high densities is that it could be of first order (see, e.g., Refs.~\cite{komoltsev:2024lcr,Ecker:2025vnb,Tang:2025xib}). If this is true, it still remains unknown whether it must be described using the Maxwell or the Gibbs thermodynamic constructions for such a type of phase transition (see, e.g., Refs.~\cite{baym:2018fht,orsaria:2019pti} and references therein). The main difference between these approaches resides in the fact that along a Gibbs phase transition, the pressure and energy density (but not the speed of sound) are continuous along the mixed phase, whereas in a Maxwell phase transition at a given transitional pressure, $P_t$, the energy density, $\epsilon$, suffers a jump, $\Delta \epsilon$. Another difference is the fact that in the mixed phase of the Gibbs construction, charge neutrality in each volume element is attained globally and not locally as in the Maxwell case. The principal quantity that determines which of the two thermodynamic constructions is favored is the hadron-quark surface tension, $\sigma_{\rm HQ}$. Values below the critical threshold, estimated to be around $\sigma_{\rm HQ}^{\rm crit} \sim 5 - 40 , \rm{MeV}/\rm{fm}^2$ favor mixed phases, whereas larger values favor a sharp interphase (see, e.g., Ref.~\cite{lugones:2021pci} and references therein). Unfortunately, this quantity is currently poorly constrained, and only a precise determination of $\sigma_{\rm HQ}$ would allow us to know the nature and properties of the hadron-quark phase transition.

Additionally, if the phase transition is sharply discontinuous, the timescale of conversion of hadrons into quarks (and vice versa) has been shown to play a key role in the dynamical stability of hybrid stars against linearized radial perturbations. In 
Ref.~\cite{Pereira:2018pte} it has been shown that if the nucleation timescale is much larger than the oscillation period of the radial perturbations, then stellar configurations in a branch where $\partial M /\partial \epsilon _c < 0$ can be dynamically stable (for a review on the subject, see, for example, Ref. \cite{lugones:2021pci} and references therein). This is the \textit{slow} conversion regime where the fluid elements in the vicinity of the transition surface keep their nature when stretched or compressed around the transition pressure, where an extended branch of \textit{slow stable hybrid stars} (SSHS) might appear. In the opposite case, the \textit{rapid} regime, the standard stability criteria ($\partial M /\partial \epsilon _c \geq 0$) remain valid\footnote{ {Some few works, e.g. Ref. \cite{Rau:2023wiq}, explored the case of intermediate timescales between rapid and slow conversions. However, precise details about the microphysics should be known, thus making the outcomes model dependent. We left this for a future work.}}. The \textit{slow} transition hypothesis has been used to study astrophysical implications in several past works \cite{mariani:2019mhs,malfatti:2020dba, tonetto:2020dgm, Rodriguez:2021hsw, curin:2021hsw, Goncalves:2022ios, Mariani:2022omh, Ranea:2022bou, lugones:2023ama, Ranea:2023auq, Ranea:2023cmr, Rau:2023tfo, Rau:2023neo, Rather:2024roo, jimenez:2024htq,Laskos:2025aph}. Such studies prove the robustness of the SSHS scenario independently of the nature of the hybrid EoS used. Besides, we found in our calculations the presence of twin, triplet, and sometimes even quadruplet stars, i.e. one, two, and three hybrid stars having the same gravitational mass but different radii than the purely hadronic sibling \cite{jimenez:2024htq}. For our SSHSs, twins, triplets, and quadruplets do not coincide with those found in the literature where only rapid-stable hybrid stars were considered \cite{Christian:2017jni}. Besides, we will see that these multiplet configurations have different discontinuous $g$-mode frequencies in a given family of SSHSs.



 {The projected Einstein Telescope (ET) and Cosmic Explorer (CE) are expected to exhibit sensitivities at least one order of magnitude better than LIGO A+ in the complete frequency range of frequencies (see, for example,  Refs. \cite{Hall:2022cea, capote:2025ald} and references therein). This significant sensitivity improvement over a broad range of frequencies in the range between $\sim 10$ Hz and a several kHz, will expand the observable universe for neutron star-neutron star (NS-NS) mergers to redshifts of up to approximately 10 (see, e.g., Refs. \cite{Hall:2022cea,Abac:2025tso} and references therein). Such an extensive observable volume could enable us to detect closely merging systems in unprecedented detail, potentially allowing for observations beyond just the fundamental mode of oscillation (see, e.g., Ref. \cite{counsell:2025imi} and references therein).} Moreover, with such an improvement in the sensitivities in a broader range of frequencies, the detection of gravitational waves emitted by isolated NSs (see, e.g., Refs.~\cite{lasky:2015gwf,haskell:2022ins, Andersson:2021agw,Font:2025gwf} and references therein) might become possible.

In this context, the study of the quasi-normal modes becomes of paramount importance. Particularly, cold hybrid stars have a distinctive $g$-mode associated with the discontinuity in the energy density that can be excited only if the conversion speed is \textit{slow} {, the so-called ``discontinuity $g$-modes\footnote{ {Usually in the literature this mode is also known as the discontinuous-density-jump interface mode, $i_\rho$ \cite{Zhu:2022pja}. Along this work, we will refer to them just as $g$-modes for simplicity.}}"} (see, e.g., Ref.~\cite{tonetto:2020dgm} and references therein). Several studies have focused attention on this particular mode \cite{sotani:2001ddo,vasquez:2014dha,ranea:2018omo,Rodriguez:2021hsw,Mariani:2022omh,zhao:2025soc_arxiv}. Moreover, universal relations associated with non-radial oscillation modes, such as the $g$-mode studied here, play a key role in asteroseismology (see the pioneering work Ref.~\cite{Andersson:1998tgw}, the recent review of Ref.~\cite{Andersson:2021agw} and references therein).

The general idea associated with universal relations including non-radial quasi-normal modes is to obtain (almost) EoS independent relationships between the frequency and damping times of particular eigenmodes and macroscopic quantities of the pulsating compact object. For example, in the case of the fundamental mode ($f$-mode), several universal relations relate their associated frequencies with the gravitational mass, $M$, and the radius, $R$ (see, e.g., Refs.~\cite{Andersson:1998tgw,benhar:2004gwa,tsui:2005uiq,chirenti:2015fom}). Other relationships relate the $f$-mode frequencies to the moment of inertia, $I$, (see, e.g., Refs.~\cite{lau:2010ipp,chirenti:2015fom}) and dimensionless tidal deformability, $\Lambda$  \cite{sotani:2021urb,pradhan:2024phq}. Of course, there exist also universal relations for the $p$-modes \cite{Andersson:1998tgw,benhar:2004gwa}, $g$-modes (associated with sharp hadron-quark phase transitions but not for long branches of SSHS  {-typically associated to large values of $\Delta \epsilon$}- as the ones of the present work) \cite{ranea:2018omo,Rodriguez:2021hsw} and also the axial and polar $wI$-modes \cite{benhar:2004gwa,tsui:2005uiq,blazquezsalcedo:2013prf,Ranea:2023cmr}.

However, in spite of the enormous amount of work that explores these quasi-normal modes in hadronic, quark, and hybrid stars (see, e.g., Ref. \cite{Guha:2024gfe}), we realized that in the current literature, no work has explored in full detail (and using a large sample of EoSs) the discontinuity $g$-modes inside SSHSs, nor have they proposed universal relations for this specific non-radial mode inside  ultra-dense branches of SSHSs. In particular, there are no studies of this mode for the four broad categories of hybrid stars representing different properties of the strong first-order phase transition, e.g. large/intermediate values for the energy density jumps, $\Delta\epsilon$, or low/large values for the deconfinement transitional pressures, $P_t$. Related to that, it is worth mentioning that although Refs.~\cite{tonetto:2020dgm,Rodriguez:2021hsw,Guha:2024gfe} explored {NSs} with first-order discontinuities in the EoS, their stellar families assumed low values of $\Delta\epsilon$ \cite{Gerlach:1968zz,Schertler:2000xq,Alvarez-Castillo:2017qki}.


For all these reasons, we believe that a careful investigation of the discontinuity $g$-modes inside hybrid stars is mandatory since it has been understood some years ago that any potentially observed hybrid star could be classified within the four categories explored in detail in Refs.~\cite{Christian:2017jni,Christian:2023hez} for which one can, in principle, build universal relations for each one of them relating stellar structural and oscillatory properties to features of the hybrid EoS. In particular, the dependency of the discontinuity $g$-mode frequencies against masses, tidal deformabilities, and radii. In fact, it was proven in Ref.~\cite{tonetto:2020dgm} that these kinds of calculations for this $g$-mode should be done only for the SSHS, i.e. slow phase conversion dynamics allow finite computations of the associated frequencies, whereas rapid phase conversions will never produce finite values of frequency for this non-radial mode.

The main goal of this work is to study the $g$-mode frequencies of HSs within the relativistic Cowling approximation for the four categories, assuming that only slow phase conversions occur at the radial interface separating the hadronic and quark phases, which in turn affect the boundary conditions of the non-radial quasi-normal mode equations. The paper is organized as follows. In Sec.~\ref{sec:theory}, we present basic theoretical aspects about the four categories of hybrid EoSs considered in this work, which are described using the constant-speed-of-sound (CSS) parametrization. Besides, we describe the non-radial perturbation equations within the relativistic Cowling approximation, adding also information about the slow junction conditions at the interface point. In Sec.~\ref{sec:res1} we present our findings for the behavior of the $g$-mode frequencies depending on the masses and tidal deformabilities. Moreover, in Sec.~\ref{sec:res2} we analyze the relations between the $g$-mode frequency and the normalized energy density jump within each category of hybrid stars. Building on this, we then propose a novel universal relation that uniquely encompasses all four categories, including configurations with long branches of slow stable hybrid stars. Finally, Sec.~\ref{sec:conc} is devoted to presenting a summary of our key findings, in particular, to discuss our proposed universal relations and their astrophysical implications when future multimessenger data become available allowing us to distinguish HSs from other types of compact stars. For completeness, radial profiles of the perturbation functions for the $g$-modes are provided in Appendix~\ref{App:A}, while Appendix~\ref{App:B} presents the formalism used for the calculation of the dimensionless tidal deformability in NSs with a first-order phase transition.
\section{Microphysics and \\quasi-normal asteroseismology} \label{sec:theory}

As already pointed out, in this section we explain the fundamental aspects of the physics underlying the EoS for hybrid star matter for the four categories, as well as how these aspects enter the non-radial oscillation equations when determining the oscillation frequencies of the $g$-mode within the relativistic Cowling approximation. Notice that although in Subsec.~\ref{subsec:eos} we work with physical units when defining the range of validity of each EoS, whereas in Subsec.~\ref{subsec:nonrad} and in the Appendices we adopt geometrical units ($G=c=1$) for simplicity.

\subsection{Hybrid equations of state} \label{subsec:eos}

\begin{figure}[htbp]
    \centering
    \includegraphics[width=0.99\linewidth]{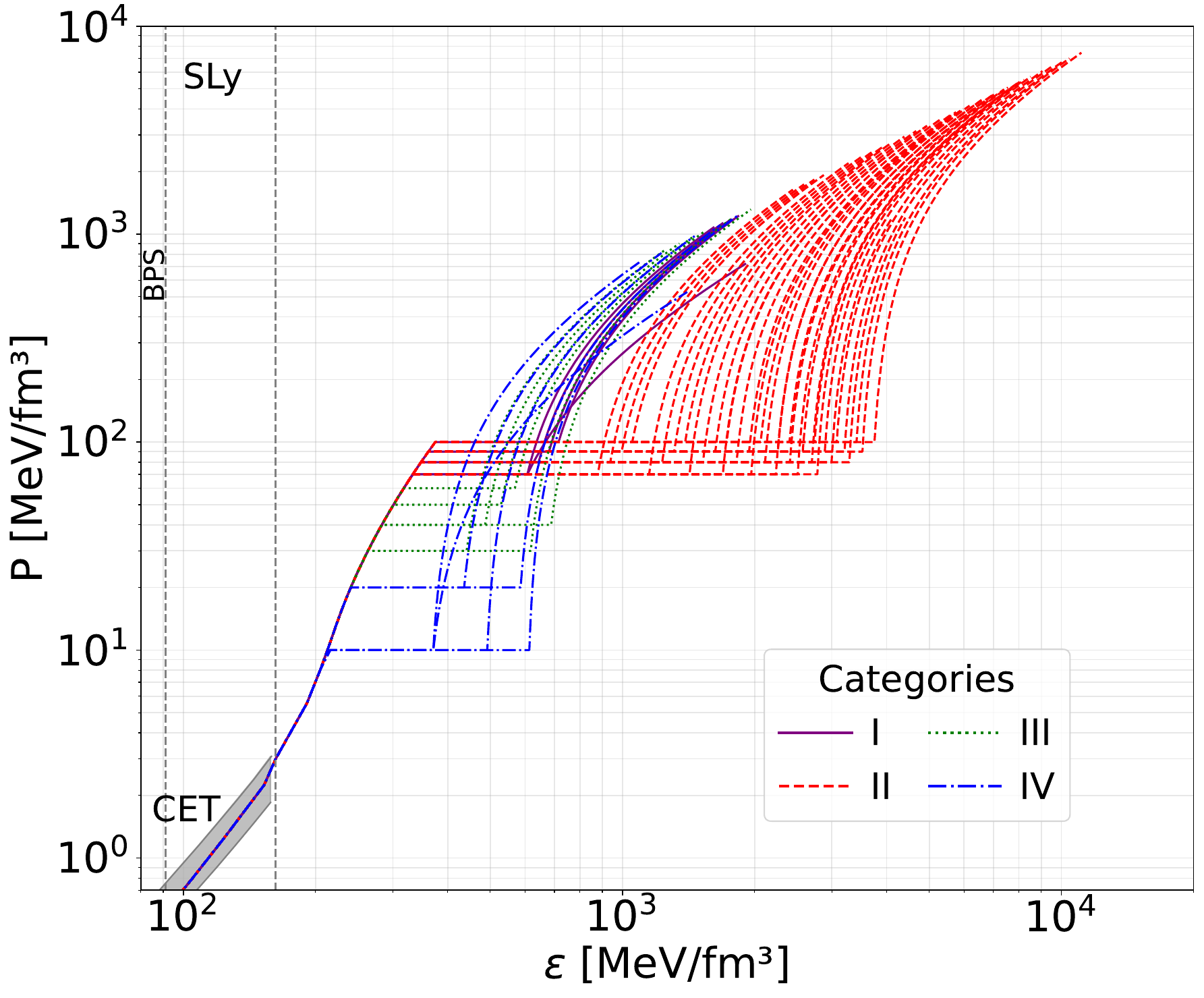}
    \caption{Set of EoSs, $P=P(\epsilon)$, for hybrid matter for the four categories explored in this work. We included the region of densities for the crust, i.e. the BPS EoS, and the exterior hadronic mantle, i.e. the SLy EoS. This last SLy EoS is then smoothly matched to the GPP and CSS EoSs at higher densities, where one can classify them in different categories (labeled I–IV) meaning different physics around the transition point. Additionally, the CET band is shown at low densities just to illustrate the agreement of our employed hybrid EoSs with this well known first-principle hadronic EoS.}
    \label{fig:eos}
\end{figure}

Since its inception some decades ago (see Ref.~\cite{Gerlach:1968zz}), the EoSs employed to obtain  hybrid star configurations considered a matching between a hadronic and a quark matter EoS. The aforementioned matching could be done by applying the Maxwell (with no mixed phase \cite{Alvarez-Castillo:2017qki}) or the Gibbs-Glendenning (with the presence of a seizable mixed phase \cite{Schertler:2000xq,Glendenning:1998ag}) constructions. However, as it is still unclear which type of construction follows the QCD deconfinement transition at finite baryonic densities, one has the freedom to choose between the two possibilities. For the present work, we choose the Maxwell construction\footnote{There are other type of $g$-modes associated with the internal composition and chemical equilibrium in the Gibbs mixed phases \citep{Wei:2018tts,Jaikumar:2021jbw,Constantinou:2021hba,Zhao:2022toc,Kumar:2021hzo} with recent results at finite temperatures in Ref. \cite{Gittins:2024oeh}.} since it allows a large parameter space  of viable hybrid stars, which satisfy the classical Seidov criterion \cite{Seidov:1971tso}. The Seidov criterion determines locally the derivative of the $M$-$R$ curve right after an abrupt phase transition. Historically, it has been used as a stability criteria to determine if after the hadron-quark phase transition, hybrid stars become unstable. Depending on the nature of the EoS used to describe quark matter, stability might be recovered giving rise to a stable branch of hybrid stars commonly referred as the third branch of stable stars (see, e.g., Ref.~\cite{Christian:2023hez}). It is important to recall that, despite determining the shape of the $M$-$R$ curve after the phase transition, when considering the slow conversion scenario between hadronic and quark matter, the Seidov criteria can not be used to determine stability against radial perturbations (see, Ref. \cite{mariani:2026sso-arXiv} for a more detailed discussion). 

As already mentioned, we use the CSS parametrization for the quark sector of the hybrid EoSs \cite{alford:2013gcf}:

\begin{equation}
  \label{eq:css_eos}
  \epsilon(P) = \begin{cases}
    \epsilon_{\mathrm{H}}(P) \hfill P < P_{t} \,, \\
    \epsilon_{\mathrm{H}}(P_t) + \Delta\epsilon + c_s^{-2}(P - P_{t}) \hspace{.5cm} P > P_{t}\,,
  \end{cases}
\end{equation}
where $\epsilon_{\rm H}$ is the hadronic EoS dependent on the pressure until the transition point $P_t$ before the quark phase starts above $P_t$. Moreover, $\Delta \epsilon$ is the jump in the energy density and $c^2_s$ is the squared speed of sound of the quark phase, which, within this parametrization, is also a free constant that, in general, should be $\gtrsim 0.5$ to favor \textit{totally stable} hybrid stars ({i.e.} hybrid configurations that are stable in both the slow and rapid conversion scenarios). It is worth to mention that although it seems that some of the values used for the speed of sound are in contradiction with the conformal limit of perturbative QCD (pQCD) at high densities, one should keep in mind that this limit can only be reached at unrealistic central baryon densities not realized even at the core of massive NSs nor in the most extreme cases of SSHSs. In this sense, it is not expected that the CSS EoS captures the conformality bound of pQCD. Related to SSHSs, this limitation of the CSS parametrization have been discussed with detail in Refs. \cite{mariani:2026sso-arXiv,zhang:2026cas-arxiv}, where alternative parametric models to describe the quark sector of hybrid EoSs are presented. Such models present non-constant values for $c_s^2$ that are in agreement with pQCD calculations.

Since all the information about the quark phase is encoded in the set of parameters $\lbrace \Delta\epsilon,~c_s^{2},~P_{t} \rbrace$, then we need further information about the hadronic EoS to determine consistently the term $\epsilon_{\mathrm{H}}(P_t)$. For this work, we use the famous Skyrme-Lyon (SLy) nuclear EoS in its tabulated version \cite{Douchin:2001sv} for baryonic densities between \mbox{$0.5792\,n_0$} and $1.1~n_0$. Notice that this is the same range of validity of the CET band of results for the stiff, intermediate and soft EoSs \cite{Hebeler:2013nza}. As one can see from Fig.~\ref{fig:eos}, the SLy EoS lies approximately in the middle of the CET band (which is included for comparison) {at low densities}, but it becomes stiffer at increasing densities,  {which we consider a more realistic approach}. This is in marked contrast to past works where only the stiffer CET EoS is employed to ensure the existence of the hybrid star family. As usual, for baryon densities below \mbox{$0.5792\,n_0$}, we employ the Baym-Pethick-Sutherland (BPS) \cite{Baym:1971tgs} EoS for the crust until the stellar surface.

After fixing the EoS for the hadronic phase up to \mbox{$n_{B}=1.1~n_0$}, we point out details of the hadronic EoS at higher densities until reaching the quark phase, as displayed in Fig.~\ref{fig:eos}. For this density sector, we use the EoSs of Sec. IIA of Ref.~\cite{jimenez:2024htq} which before the phase transition employs three\footnote{ {Our GPP parameters were adjusted to a) agree with the stiff CET EoS at $1.1n_0$, b) respect the causality limit, and c) respect the $2M_\odot$-mass constraint. See Refs. \cite{jimenez:2024htq,Lugones:2021bkm} for further details.}} model-agnostic generalized piecewise polytropes (GPP) introduced in Ref. \cite{OBoyle:2020peo}. Notice that these GPP's were also introduced to built thermodynamic-consistent EoSs by enforcing continuity not only of the pressure and energy density (as usually done in the literature), but also the speed of sound. At the transition point, we made use of the same CSS parameters $\lbrace \Delta\epsilon,~c_s^{2}\sim [0.5,1],~P_{t} \rbrace$ as those from Ref.~\cite{jimenez:2024htq} for the studied SSHS in that work.

It should be noted that in Ref.~\cite{jimenez:2024htq}, SSHSs were considered only in very specific cases, mainly for comparison with the rapid conversion scenario. Although SSHSs share similar equations of state, they are more numerous, since they can exist in regions where rapid-conversion HSs become unstable. In contrast, in the present study we restrict ourselves to the slow conversion regime, as discontinuity $g$-modes are excited exclusively under these conditions, and because Ref.~\cite{jimenez:2024htq} has already demonstrated that SSHSs in the four categories of hybrid stars are dynamically stable against radial oscillations when slow junction conditions are applied. 

 As usually done in the literature, we follow the nowadays standard classification of hybrid stars of four categories depending on the gravitational masses at which the phase transition occurs in the mass-radius diagram \cite{Christian:2017jni,Montana:2018bkb,Christian:2018jyd,Christian:2020xwz,jimenez:2024htq}. The classification for each category must satisfy the following astrophysical features: i) Category I with the maximum masses of the hadronic and HS branches surpassing the $2~M_{\odot}$ limit; ii) Category II with the maximum mass of the hadronic branch reaching the $2~M_{\odot}$ value; {iii) }Category III with the maximum of the hadronic family lying between 1 and 2 $M_\odot$, and the HS star branch surpassing the $2~M_{\odot}$ limit; and iv) Category IV with the maximum of the hadronic branch lying below 1 $M_\odot$ while the HS branch surpasses the $2~M_{\odot}$ limit.  {See Ref. \cite{Montana:2018bkb} for further details on the causes for the behavior of each branch in each category.} For completeness, in Table \ref{tableEoSs}, we list a range of values used for each transitional quantity. Within each range, one can choose the extreme values corresponding to representative EoS. For instance, the minimal values in $\Delta \epsilon$ correspond to the minimal values of $P_t$ and so on. It was verified in Ref. \cite{jimenez:2024htq} that each range produces exactly the four categories of hybrid stars of Refs. \cite{Christian:2017jni,Montana:2018bkb}.

 A comment is in order concerning the chosen values for the parameters of our quark EoSs. At first sight, it may seem that some of our CSS parameters of a given category are extreme compared to some constraints  {obtained through Bayesian inference techniques using some modern astrophysical data. For instance, Ref. \cite{brandes:2024cop} constraints the values of $\Delta\epsilon/\epsilon_t$ to be less than 0.2, while several of our hybrid stars require this quantity to be larger than 1. On one hand, it should be noticed that even some of the standard approaches constraining phase-transition parameters (see, e.g., Refs. \cite{Christian:2023hez,Li:2024lmd,Blacker:2024tet}) conclude that it is too early to rule out twin stars with current astrophysics data. Only very precise measurements of some specific observables (like radii differences between a NS and its twin) as well as oscillatory properties (like non-radial modes) in the future data will help us to have a reliable answer. On the other hand, Bayesian and other agnostic studies disfavoring strong transitions} arrive to this conclusion mainly because they do not explicitly include the slow conversion possibility that produces SSHSs.  {If taken into account, these approaches could yield $\Delta\epsilon/\epsilon_t \gg 0.2$. For instance,} despite not being free of controversy, the recent mass $M = (1.21 \pm 0.05\,) M_\odot$ and radius $R = (7.0 \pm 0.4)$ estimation of the compact object XTE J1814-338 \cite{kini:2024ctp} put tension to traditional ideas about quark matter at NS cores.  {Interestingly, SSHBs with $\Delta\epsilon/\epsilon_t >5$ were proven to naturally produce objects in like XTE J1814-338 \cite{Laskos:2025aph,mariani:2025hns}. In this sense, slow conversions with large $\Delta \epsilon/ \epsilon_t$ are able to produce a reasonable explanation for the existence of some exotic astronomical observations already known and to be discovered (see, e.g., Ref. \cite{mariani:2025hns}). By virtue of these reasons,} we expand the traditional parameter range and add more extreme scenarios still consistent with NS data.

Interestingly, one can easily verify, using the particular EoS parameters given in Table \ref{tableEoSs}, that, as stated in Ref.~\cite{Christian:2017jni}, the maximum values of masses reached by each branch of each category directly depend on the values taken by the triad $\lbrace \Delta\epsilon,~c_s^{2},~P_{t} \rbrace$.

\begin{table}
\begin{tabular}{| c | c | c | c | c |}
\hline 
Category & $\Delta \epsilon$ & $P_t$       & $c^2_s$   & $\epsilon_t$  \\ \hline \hline 
I        & $274 - 340$       &  $70 - 100$ & $0.5 - 1$ & $333 - 374$   \\
II       & $545 - 3376$      &  $70 - 100$ & $0.5 - 1$ & $333 - 374$   \\  
III      & $178 - 405$       &  $30 - 60$  & $0.5 - 1$ & $263 - 317$   \\ 
IV       & $154 - 397$       &  $10 - 20$  & $0.5 - 1$     & $215 - 240$   \\ \hline   
\end{tabular}
\caption{{Selected ranges of values for the EoS parameters for each category, where} $\Delta \epsilon$ denotes the jump in energy density at the transition; $P_t$ is the transition pressure; $c_s^2$ is the squared speed of sound, normalized to the speed of light; and $\epsilon_t$ is the energy density at the transition. All quantities, except $c_s^2$, are expressed in units of MeV/fm$^3$.}\label{tableEoSs}
\end{table}

\subsection{Non-radial modes in the Cowling approximation} \label{subsec:nonrad}

As already summarized in the Introduction, our main aim of the present work is to calculate the frequencies associated with the $g$-modes produced by discontinuities in the EoS coming from first-order phase transitions, which in our case are manifested as hybrid stars. For this, we assume that these ultra-dense compact objects possess spherical symmetry, i.e. no uniform nor differential rotation. 
This hypothesis allows us to use the Tolman-Oppenheimer-Volkov (TOV) equations to determine their structural features, such as masses and radii, {and thereby construct} the corresponding mass-radius relation to be compared with modern astronomical data.

{I}n general, the calculation of the non-radial mode frequencies is done by studying quadrupolar perturbations 
considering time-dependent even-parity perturbations of the metric functions coupled to perturbations of the thermodynamic variables. This is the so-called general-relativistic linear perturbation theory \cite{tonetto:2020dgm}. Unfortunately, the numerical cost to obtain these results is quite expensive, and one usually uses a reasonable approximation called the `relativistic Cowling approximation' \cite{finn:1988rsp}, where only the fluid perturbations are considered in a fixed background space-time. Interestingly, this simplification gives quite accurate results for the oscillation frequencies when compared to the full problem (see, e.g., Refs.~\cite{sotani:2011soh,chirenti:2015fom} and references therein). In the particular case of the $g$-modes, the differences between both calculations are only of $\sim 10\%$ \cite{sotani:2001ddo,tonetto:2020dgm}. Unfortunately, one of the major drawbacks of this approximation is that the damping times of these quasi-normal modes cannot be obtained. In any case, {given the proven reliability of this approach for determining oscillation} frequencies, we also use the Cowling approximation when {calculating} the $g$-mode of SSHS. Our results capture both the qualitative and quantitative features of the full (numerically challenging) calculation, which, in some cases, may become unreliable and still require additional simplifications to clearly distinguish the $g$-mode from other non-radial modes with comparable values \cite{tonetto:2020dgm}.


Within this approximation, we can still work with the spherically-symmetric spacetime having a line element that can be written as
\begin{eqnarray}
    {\rm d}s^2 = -e^{\nu (r)} {\rm d}t^2 + \left( 1- \frac{2m(r)}{r}\right)^{-1} {\rm d} r^2 +
    && r^2 d\Omega^2 \, ,
\end{eqnarray}
where $d\Omega^2 = {\rm d}\theta ^2 + \sin ^2 \theta {\rm d}\phi ^2$ is the metric on the two-sphere,  $\nu (r)$ and $m(r)$ are the metric and mass profiles, respectively, obtained after solving the TOV equations (with appropriate boundary conditions). Notice that although the TOV equations provide solutions usually for the pressure or energy density profiles, $P(r)$ or $\epsilon(r)$, one can also obtain $\nu(r)$ from these solutions \cite{glendenning2012compact}. 

{It is worth mentioning that the onset of strong transitions is {(generally but not always)} manifested as kinks in the $M$-$R$ relations, then exhibiting decreasing values of $\lbrace M,~R\rbrace$ for increasing central pressures ($P_{\rm c}$). From the hydrostatic perspective (and rapid phase-conversion viewpoint), these $M$-$R$ sectors {close to the phase transition} are {often} considered unstable since the values of central pressure are small compared to the magnitude of $\Delta\epsilon$, thus having values of $P_{\rm c}$ not enough to counterbalance such high values of corresponding central energy density. For instance, in Fig.~\ref{fig:eos}, Category II EoSs have $\Delta\epsilon\sim 3000~{\rm MeV/fm^3}$, while central pressures above the transition are $P_{\rm c}\sim 200~{\rm MeV/fm^3}$.} 



Following the usual procedure presented in Ref.~\cite{sotani:2011soh}, after decomposing the perturbation equations in spherical harmonics while assuming a harmonic time dependence ($\sim e^{i\omega t}$ with $\omega$ being the non-radial oscillation frequency), the relativistic Cowling equations governing the behavior of the fluid perturbations (characterized by functions $V$ and $W$) that need to be solved to obtain the non-radial mode oscillation frequencies are:
\begin{eqnarray}\label{eqn:dwdr}
    \frac{{\rm d}W(r)}{{\rm d}r} &=& 
    \frac{{\rm d}\epsilon}{{\rm d}P} \Bigg[\omega^2\frac{r^{2}}{\sqrt{1-\frac{2m(r)}{r}}}e^{-\nu(r)} V(r)  + \\ \nonumber
    &&\frac{1}{2}\frac{{\rm d}\nu(r)}{{\rm d}r} W(r)\Bigg]\ - \frac{\ell(\ell + 1)}{\sqrt{1-\frac{2m(r)}{r}}} V (r)
\end{eqnarray}{}
{and}
\begin{equation}\label{eqn:dvdr}
    \frac{{\rm d}V (r)}{{\rm d}r} = \frac{{\rm d}\nu(r)}{{\rm d}r} V (r) - \frac{1}{\sqrt{r-2m(r)}}W(r)  \ .
\end{equation}

In this manner the spatial part of the Lagrangian perturbation, $\xi_i$, is characterized as \cite{ranea:2018omo}:

\begin{equation*}
\xi_i = \left(\sqrt{1-\frac{2m(r)}{r}}W,-V\partial_\theta,-V\sin^{-2}\theta \partial_\phi \right)r^2Y_{\ell m}.
\end{equation*}

In order to solve Eqs.~\eqref{eqn:dwdr}-\eqref{eqn:dvdr}, we {choose $\ell=2$ (quadrupole oscillations) and} impose boundary conditions at the stellar center ($r \sim 0$), where the behavior of the solutions is \cite{sotani:2011soh}
\begin{equation}\label{eqn:borde1}
    W(r) \sim r^{\ell+1}  \ , \  V (r) \sim -\ell^{-1}r^{\ell} \ ,
\end{equation}
as well as at the stellar surface ($r = R$) where the Lagrangian perturbation to the pressure needs to vanish,
\begin{equation}\label{eqn:borde2}
    \Delta P(R) = 0 \ .
\end{equation}
This last condition can be written as
\begin{equation}\label{eqn:borde}
    \omega^{2}\left[{\sqrt{1-\frac{2M}{R}}}\right]^{-\frac{3}{2}}V(R) + \frac{1}{R^{2}} W(R)=0  \ .
\end{equation}
The values of $\omega ^2$ that satisfy the boundary value problem given by the differential Eqs.~\eqref{eqn:dwdr}-\eqref{eqn:dvdr} and the boundary conditions in Eqs. \eqref{eqn:borde1}-\eqref{eqn:borde} are the frequencies of the non-radial modes we are interested in obtaining.

For our case of hybrid stars having an EoS containing a discontinuity (characterized by the $\Delta \epsilon$ term) located at a particular radial coordinate, $r = r_t$, additional junction conditions at the discontinuity surface need to be imposed \cite{vasquez:2014dha,ranea:2018omo}: 
\begin{equation}\label{eqn:pegado3}
    W_{+}=W_{-} \ ,
\end{equation}
\begin{eqnarray}\label{eqn:pegado4}
    V_{+} &=& \frac{e^{\nu}}{\omega ^2 {r_t}^{2}} \left(1 - \frac{2m}{r_t}\right) \times \nonumber \\
    &&\Bigg( \frac{\epsilon_{-}+P}{\epsilon_{+}+P}\bigg[\omega^{2}{r_{t}}^{2}e^{-\nu}\left(1-\frac{2m}{r_t}\right)^{-1}V_{-} + \\ 
    &&\nu' W_{-}\bigg] -\nu' W_{+}\Bigg) \ , \nonumber
\end{eqnarray}
where the subindex $-$ ($+$) is used to characterize quantities before (after) the phase transition.  For example, for the perturbing function $W$, we have $W_-=W(r_t-0)$ and $W_+=W(r_t+0)$. The same notation has been used for the other quantities. It is important to note that, in the absence of discontinuities, the function $V(r)$ becomes continuous and  
$V_+ = V_-$.

\begin{figure*}
    \centering
    \includegraphics[width=0.5\linewidth]{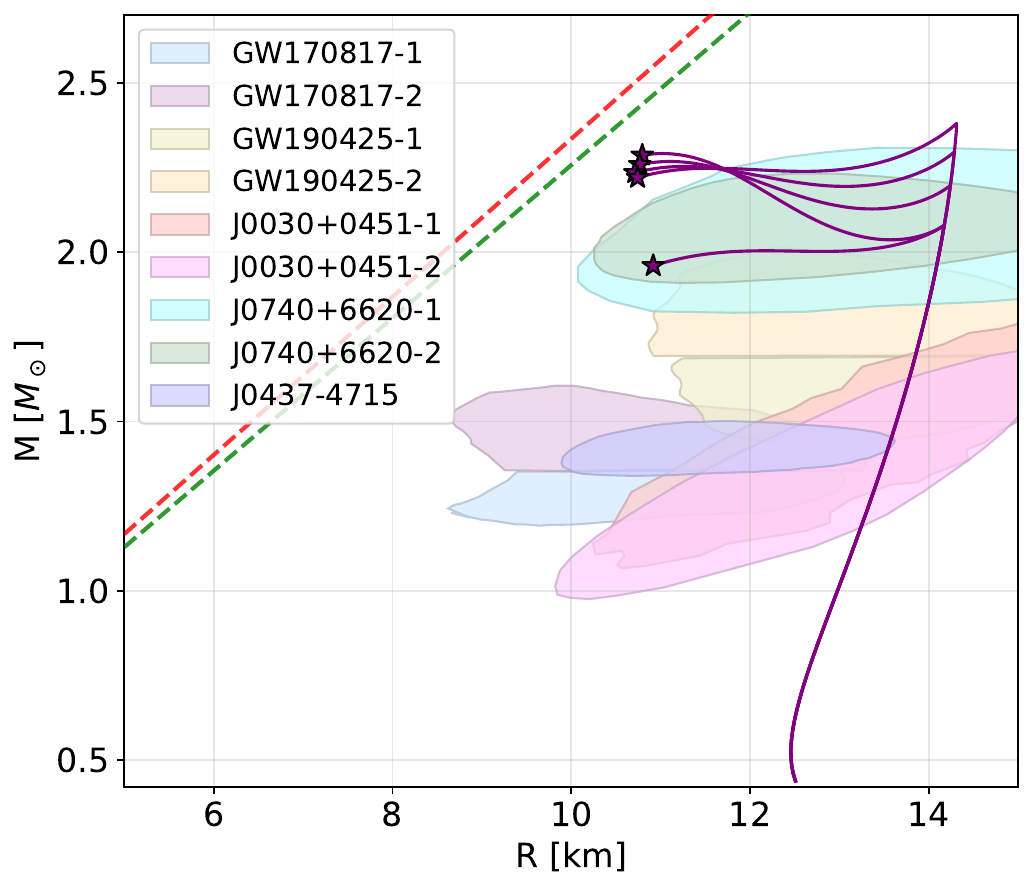}
    \includegraphics[width=1.0\linewidth]{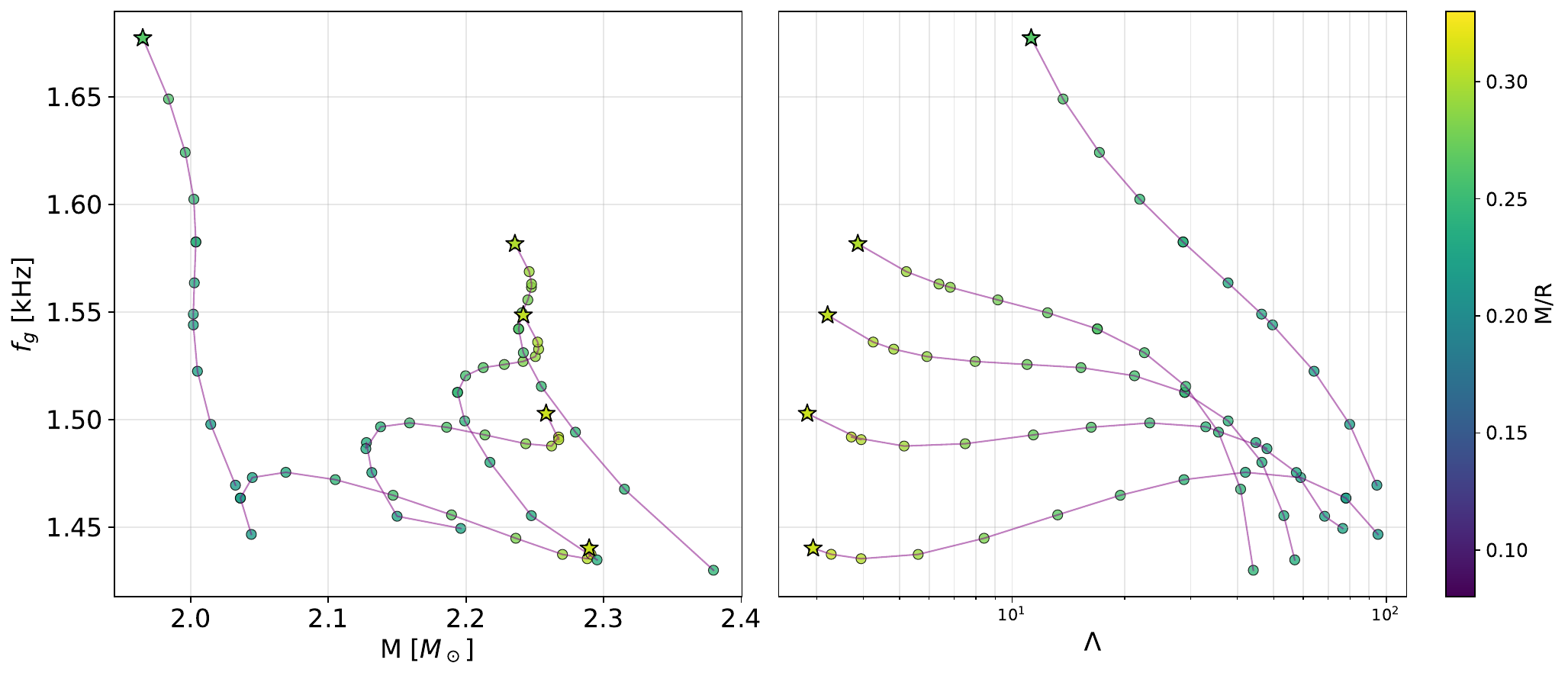}
    \caption{Results for Category I hybrid stars. {\it Top panel:} Mass–radius relations ($M$-$R$) for hadronic and hybrid stars with observational constraints from NICER (PSR J0437-4715, PSR J0740+6620, and PSR J0030+0451) and gravitational wave events (GW190425 and GW170817) for comparison.  {The red dashed diagonal line corresponds to the widely known theoretical bound on maximal compactness of $M/R = 1/2.9$ proposed by Lattimer and Prakash \cite{Lattimer:2006xb} while the green dashed diagonal line corresponds to a new bound of $M/R = 1/3$ recently proposed by Rezzolla and Ecker \cite{Rezzolla:2025pft} (see main text for details)}. {\it Bottom-left panel:} $g$-mode frequency, $f_g$, as a function of stellar mass, $M$, where the color gradient represents the compactness ($M/R$) of the star. {\it Bottom-right panel:} $f_g$ as a function of the dimensionless tidal deformability, $\Lambda$. 
    }
    \label{fig:categoryI}
\end{figure*}

\begin{figure*}
    \centering
    \includegraphics[width=0.5\linewidth]{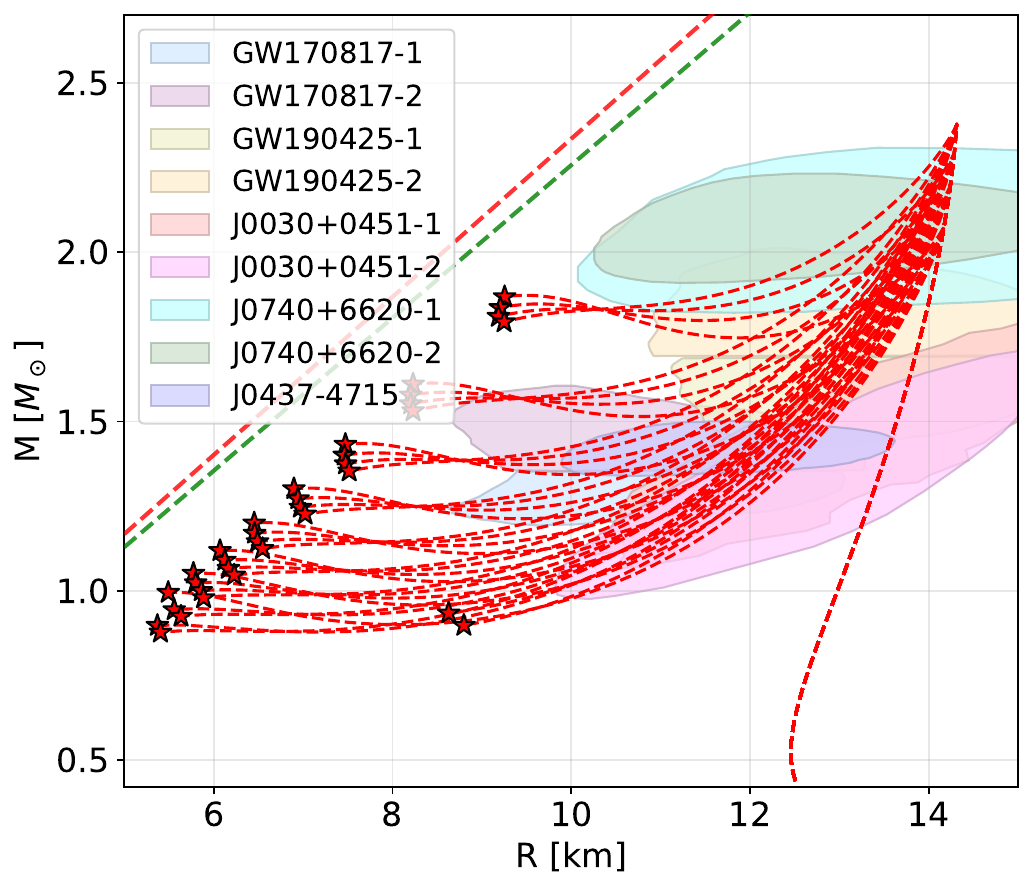}
    \includegraphics[width=1.0\linewidth]{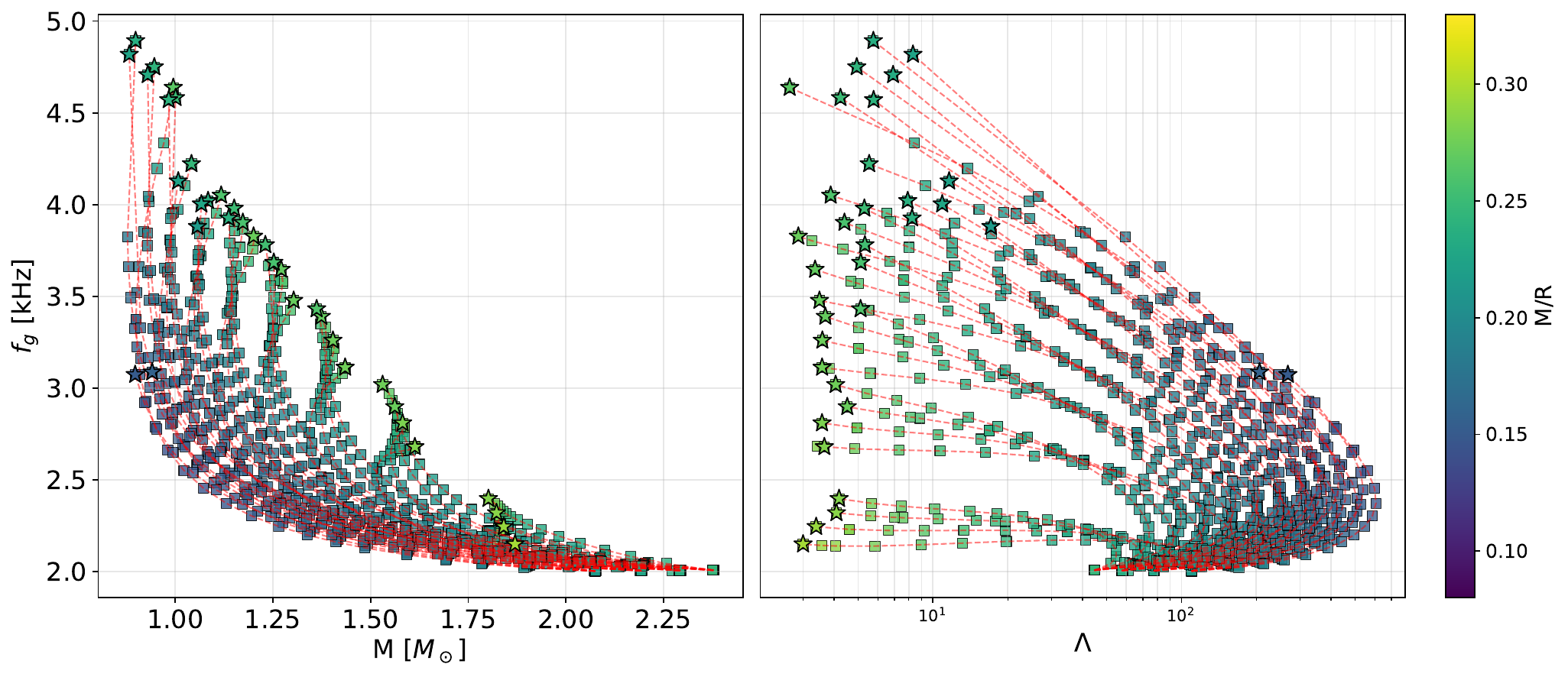}
    \caption{Results for Category II hybrid stars. Same legends and astrophysical constraints as that of Fig. \ref{fig:categoryI}. 
    }
    \label{fig:categoryII}
\end{figure*}


The frequencies of the discontinuous $g$-modes are obtained using the CFK code developed in Ref. \cite{ranea:2018omo} which solves (given a trial frequency of a given mode $n$, $\omega_{n, \rm trial}^2$) the system of Eqs.~\eqref{eqn:dwdr}-\eqref{eqn:dvdr} using a Runge-Kutta-Fehlberg integrator. Integration is performed from the center to the transition interface, where the conditions given by Eqs.~\eqref{eqn:pegado3}-\eqref{eqn:pegado4} are employed. Then, integration is performed up to the surface of the star, where the boundary condition given by Eq.~\eqref{eqn:borde2} needs to be fulfilled. To solve the eigenfrequency problem 
we apply the root-finding algorithm which is composed of a Newton-Raphson scheme coupled with Ridders’ method (see Ref. \cite{ranea:2018omo} for more details on the numerical scheme used to solve this eigenvalue problem). Furthermore, the identification of these modes as $g$-modes is supported by the analysis of the radial profiles of the eigenfunctions $W$ and $V$, as shown in Appendix~\ref{App:A}, where each eigenfunction exhibits a single node and the frequencies are found to be lower than the $f$-mode, consistent with the Cowling classification scheme.

\section{Discontinuity \texorpdfstring{$g$}{g}-modes \\ of {hybrid} neutron stars} \label{sec:res1}

\begin{figure*}
    \centering
    \includegraphics[width=0.5\linewidth]{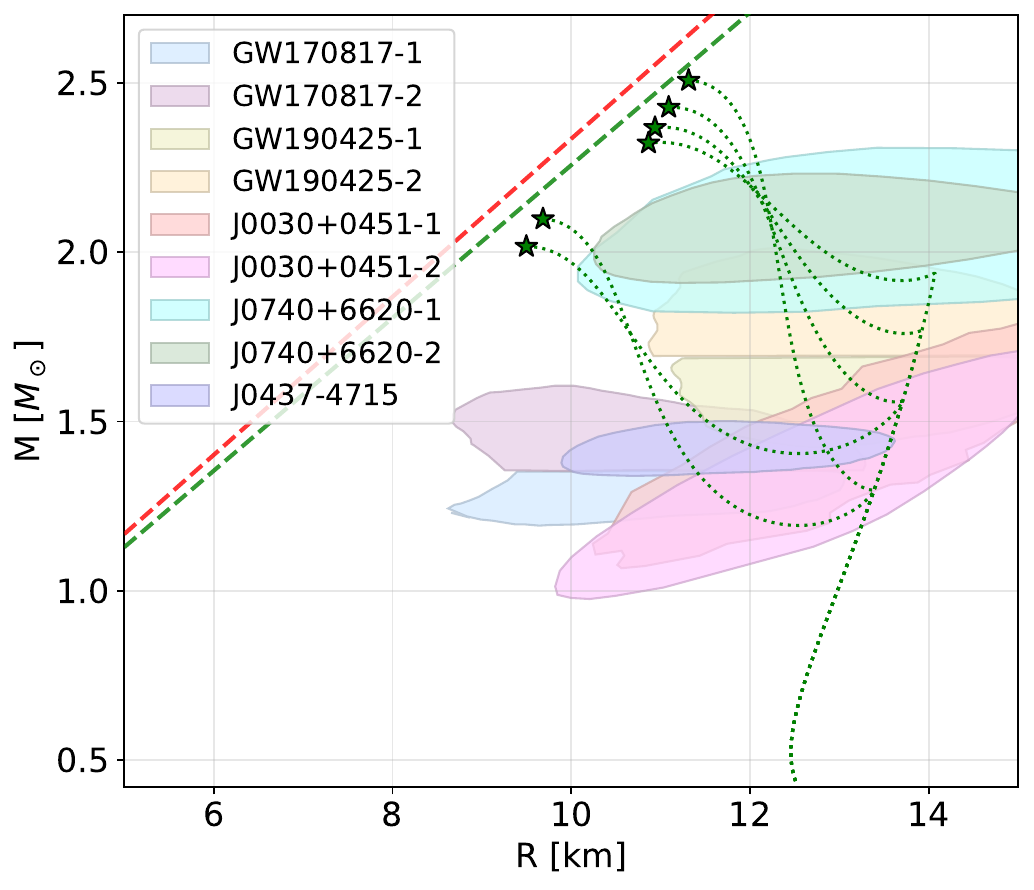}
    \includegraphics[width=1.0\linewidth]{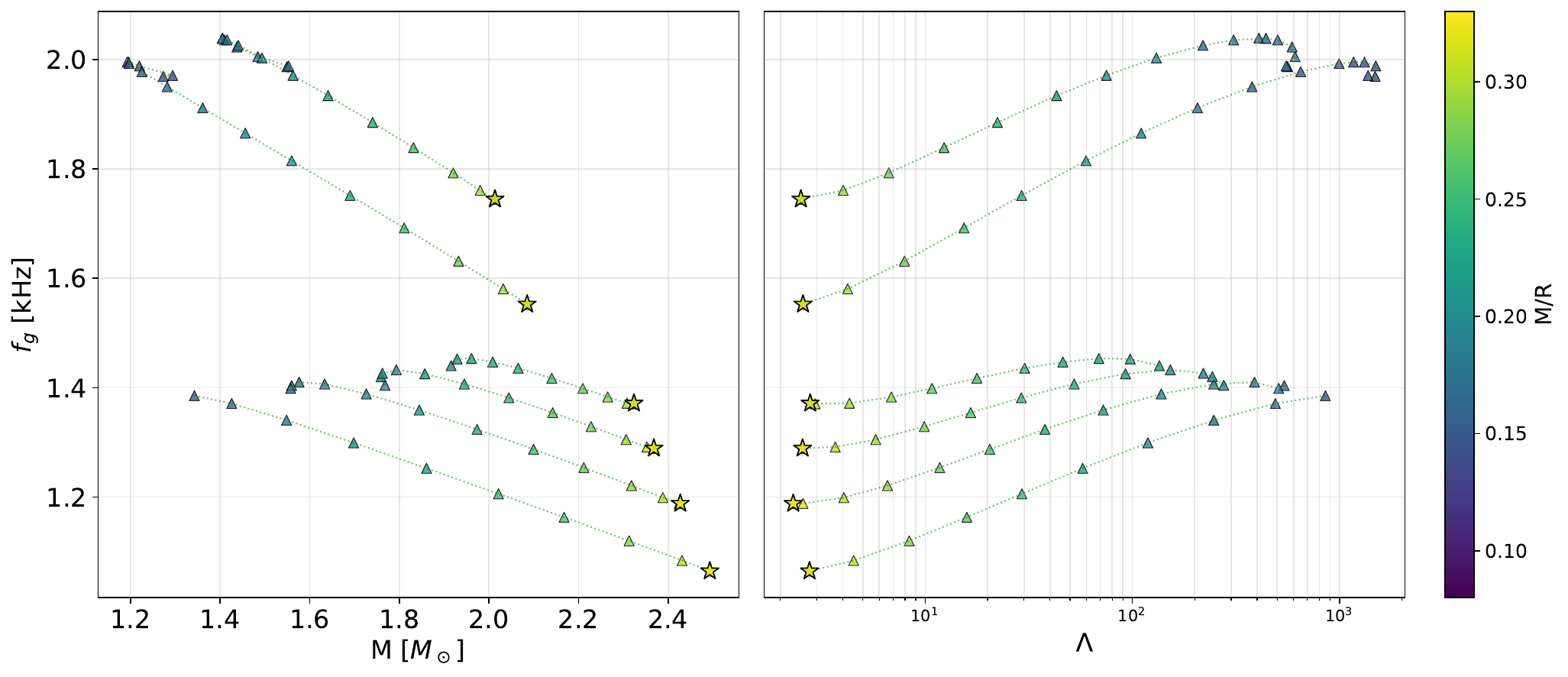}
    \caption{Results for Category III hybrid stars. Same legends and astrophysical constraints as that of Fig.~\ref{fig:categoryI}. 
    }
    \label{fig:categoryIII}
\end{figure*}

\begin{figure*}
    \centering
    \includegraphics[width=0.5\linewidth]{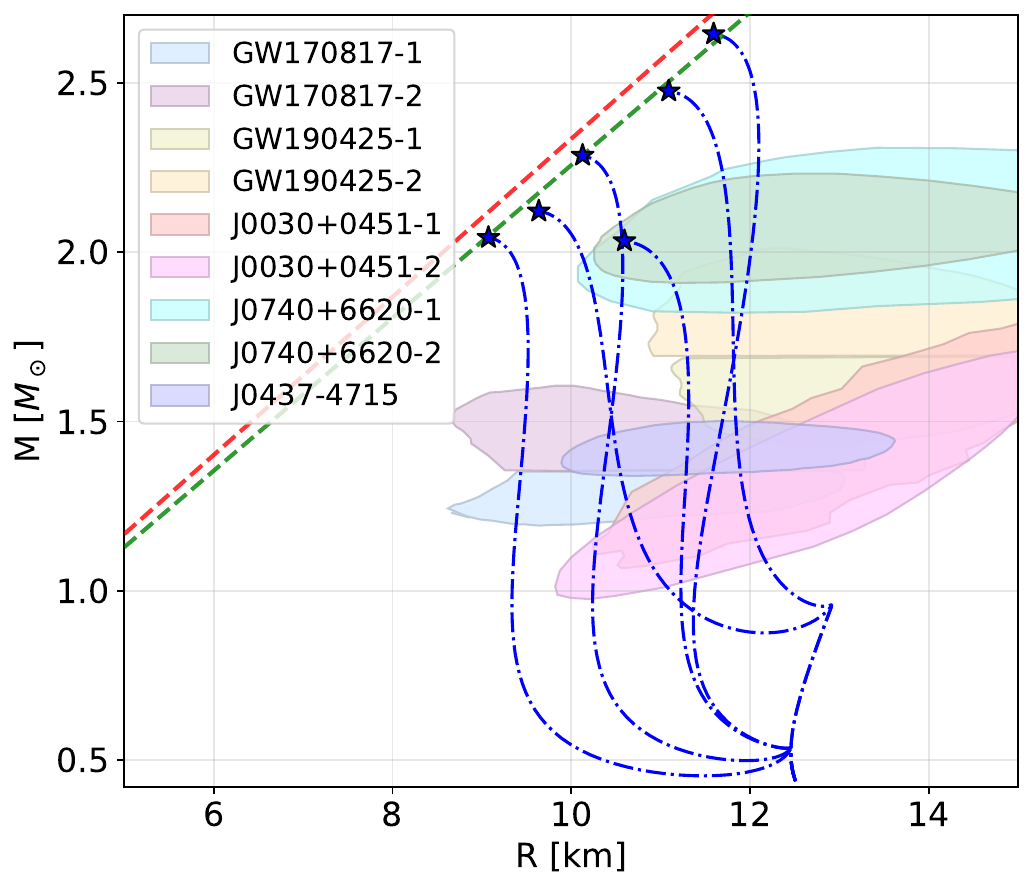}
    \includegraphics[width=1.0\linewidth]{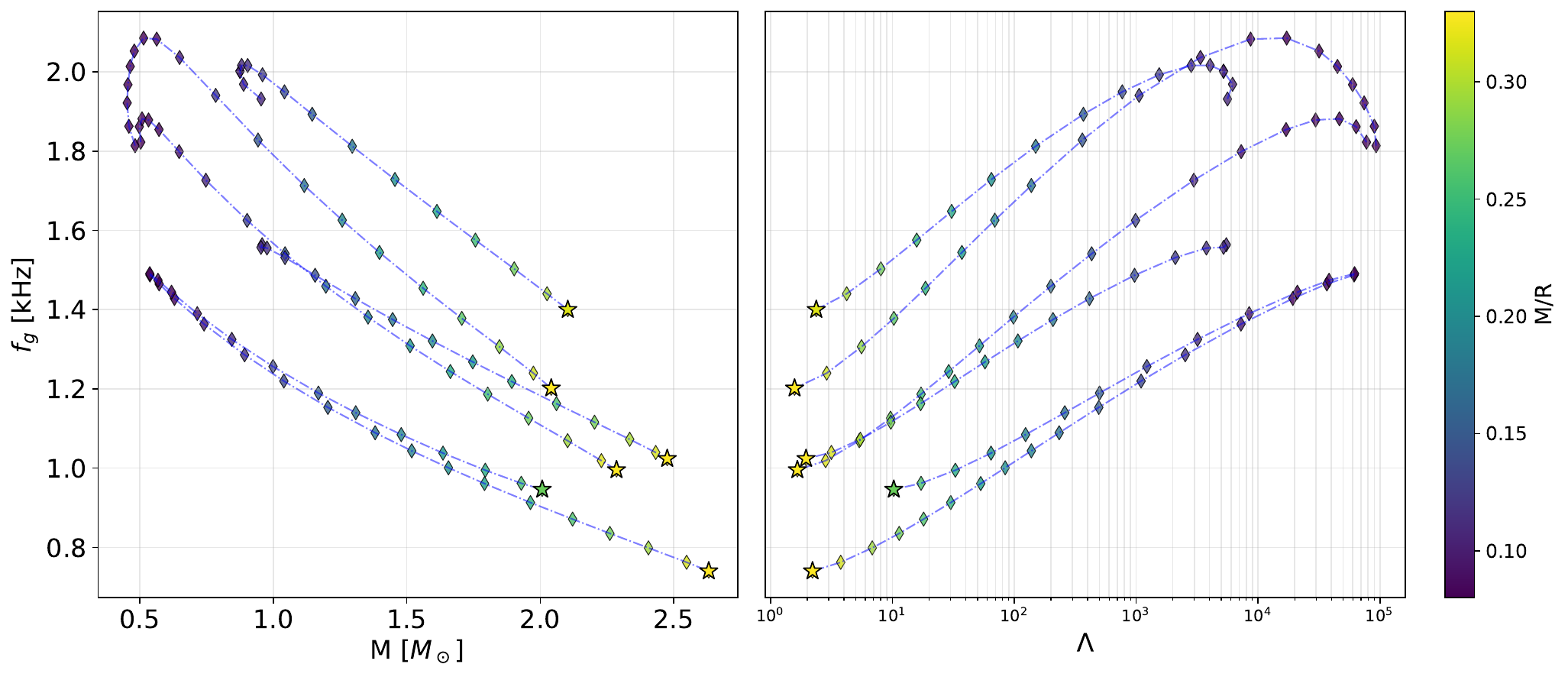}
    \caption{Results for Category IV hybrid stars. Same legends and astrophysical constraints as that of Fig.~\ref{fig:categoryI}.
    }
    \label{fig:categoryIV}
\end{figure*}
In this section, we display our results for the $g$-mode oscillation frequencies of NSs containing a strong first-order phase transition (dubbed hybrid stars) using four categories of EoSs for their interiors, as described in Section~\ref{subsec:eos}. Notice that in the top panels of Figs.~\ref{fig:categoryI}-\ref{fig:categoryIV}, we show the $M$–$R$ relations for a particular set of EoSs within each category, along with observational constraints from the binary NS mergers GW170817 \citep{Abbott:2017ogw, Abbott:2018gmo} and GW190425 \citep{Abbott:2020goo}, as well as from NICER measurements for PSR J0740+6620 \citep{Miller:2021tro, Riley:2021anv}, PSR J0030+0451 \citep{Miller:2019pjm, Riley:2019anv}, and PSR J0437–4715 \citep{choudhury:2024anv}. Besides, the legends of the astrophysical constraints in the $M$-$R$ diagram, the labels  `1' and  `2' refer to the constraints derived from Ref.~\cite{Miller:2021tro} and Ref.~\cite{Riley:2021anv}, respectively. Notice that we also display two dashed diagonal lines: the red one indicates the famous limit on maximal compactness proposed by Lattimer and Prakash \cite{Lattimer:2006xb} of $M/R=1/2.9$, where they used a maximally stiff EoS at the causal limit of $c^2_s=1$ at high densities matching it with a nuclear model at low densities. On the other hand, the green one comes from the somewhat recent work of Rezzolla and Ecker \cite{Rezzolla:2025pft} proposing the new bound $M/R=1/3$ within a more realistic approach by performing  interpolations (constrained by current astrophysics data) between CET and pQCD in a large parameter space satisfying general requirements as thermodynamic consistency and causality bounds on the speed of sound. It is worth to notice that for this Rezzolla-Ecker bound, only smooth crossovers and ultra-low-$\Delta\epsilon$ phase transitions were considered. Interestingly, our hybrid stars accounting for strong transitions seem to still satisfy this 1/3-bound, excluding some few very-massive configurations in Category IV which could be dismissed if astrophysics data is explicitly imposed at such high NS masses.

Moreover, we stress that all our curves of results for these Figs.~\ref{fig:categoryI}-\ref{fig:categoryIV} represent dynamically stable configurations of hadronic and hybrid branches due to a detailed proof given in Ref.~\cite{jimenez:2024htq}, where slow conversion dynamics and corresponding junction conditions at the transition point were assumed.  {As already stated, this choice of hypothesis  {(theorized within several microscopic models) for the phase-transition dynamics} is central, as discontinuity $g-$modes related to hadron-quark transitions can only be excited if such phase transition is sharp and the conversion speed between phases is slow.}

The bottom-left panels of Figs.~\ref{fig:categoryI}-\ref{fig:categoryIV} display the frequency of the $g$-mode, $f_g=\omega_g/2\pi$, as a function of the gravitational mass $M$, with a color gradient representing the compactness, $M/R$, of the star, also illustrating its influence on the $g$-mode frequencies. As said in the Introduction, multiplet configurations will not be restricted to just appear in the $M$-$R$ relations, but also in the $f_g$-$M$ relations where, for a given $M$, one can distinguish twin and triplets hybrid stars (sometimes even quadruplet stars in Category I) having different $f_g$.

The bottom-right panels show $f_g$ as a function of the dimensionless tidal deformability $\Lambda$. The star symbols in the panels indicate the terminal configuration, and in the bottom panels, their color further encodes the corresponding compactness.
Results among different categories are displayed using different colors and line styles, indicating the corresponding EoS category. A global comparison among the results of these four categories reveals distinct trends in the behavior of the $g$-mode frequencies, which are closely related to the properties of the transition point, e.g., large discontinuities in the energy density at the phase transition or earlier transitions to deconfined quark matter. With this in mind, one could expect quantitative and qualitative variations of $f_g$ between categories at increasing densities after the transition occurs. Moreover, it will be easy to note that all our hybrid EoS produce stellar configurations consistent with current observational constraints from NICER and gravitational-wave data.

In the following, we present the analysis of the $g$-mode frequencies for each EoS category (displayed in Figs.~\ref{fig:categoryI}–\ref{fig:categoryIV}) as follows: 

\begin{itemize}
    \item {\bf Category I:} The results related to this category are presented in Figure~\ref{fig:categoryI}. This family of hybrid stars (mostly twin but also triplet in {our present} slow conversion scenario) have frequencies, $f_g$, lying in the range \mbox{$\sim 1.4 \; - \sim 1.7$~kHz}. The SSHS near the terminal configuration have cores composed by quarks with a size of about $70\%$ of the  {radius of the} star and with a compactness around $M/R \sim 0.25$. We can see that there is no generic behavior of the frequency $f_g$ with the mass, despite in most of the cases analyzed, $f_g$ increases as the terminal configuration is reached. A similar situation occurs when the behavior of $f_g$ with the dimensionless tidal deformability, $\Lambda$, is analyzed. In general, as $\Lambda$ decreases towards the terminal configuration, $f_g$ increases. Despite this behavior can be thought to be generic, there are exceptions with some curves displaying non-monotonic behavior.

    \item {\bf Category II:} The results related to this category are presented in Figure~\ref{fig:categoryII}. The numerical values of frequencies $f_g$ within this category are notably higher than those of Category I and all other categories (see also Fig.~\ref{fig:frec-masa-lambda} below), i.e. they lie in the range $\sim 2-\sim 5$~kHz. This is unexpected since typical values for ordinary hybrid stars - stable under both the slow and rapid conversion scenarios - remain below $\sim 2.5$~kHz (see, e.g., Refs.~\cite{sotani:2001ddo,vasquez:2014dha,ranea:2018omo, guha:2025nro} and references therein). This category corresponds to the most elongated extended branches of SSHS. Nevertheless, near the terminal configuration the quark cores generally do not exceed $\sim 65\%$ of the stellar radius, and the compactness typically remains below $\sim 0.25$. For objects in this category, it becomes more evident that the frequency $f_g$ increases along the extended stability branch toward the terminal configuration, reaching higher values for lower-mass SSHS.

    \item {\bf Category III:} As can be seen from Fig.~\ref{fig:categoryIII}, the values of $f_g$ are between \mbox{$\sim 1- \sim 2$ kHz}. We separated in the $M$-$R$ diagram two families with higher and lower hybrid-star masses, i.e. $M\sim 2.4 M_{\odot}$ and $M\sim 2M_{\odot}$, respectively. We do this for clarity in the presentation of our results and analysis. With that in mind, one can see in the $f_g$-$M$ plane that the four families that reach gravitational masses $M\sim 2.4 M_{\odot}$ have lower $f_g$ compared to the two lower-mass ($M\sim 2M_{\odot}$) families.

    For this category, the general behavior of $f_g$ is that it decreases with the mass of the HS. Moreover, for this families, no long branches of SSHS appear after the maximum mass configuration where the quark core occupies more that $80\,\%$ of the star. Despite this fact, it important to remark that triplets are possible as SSHS that connect the hadronic branch with the {\it{totally stable}}\footnote{Hybrid configurations that are stable both in the slow and rapid regime of conversion between phases.} branch of HS appear after the phase transition takes place. As a function of the dimensionless tidal deformability $\Lambda$, the frequency $f_g$ decreases almost monotonically: the more compact objects in this category exhibit both lower $\Lambda$ and lower $f_g$ than the first HSs formed after the phase transition.

    \item {\bf Category IV:} As can be seen from Fig.~\ref{fig:categoryIV}, the $g$-mode frequencies lie in the range $0.7-2.1$~kHz. Despite having transition pressures lower that those of Category III, the general behavior of $f_g$ with the mass and dimensionless tidal deformability are qualitatively similar. The larger deviations between both categories can be seen for extremely low-mass HSs where $f_g$ decreases with the mass but increases with $\Lambda$ until the behavior described previously towards the terminal configuration (where compactness is larger than 0.3 and the quark core represents $\sim 85\,\%$ of the stellar configuration) is regained. As in Category II, triplets of low mass ($M < 1\,M_\odot$) are possible but no SSHS appear after the maximum mass configuration is reached.
    
    
\end{itemize}

\begin{figure}
    \centering
    \includegraphics[width=0.99\linewidth]{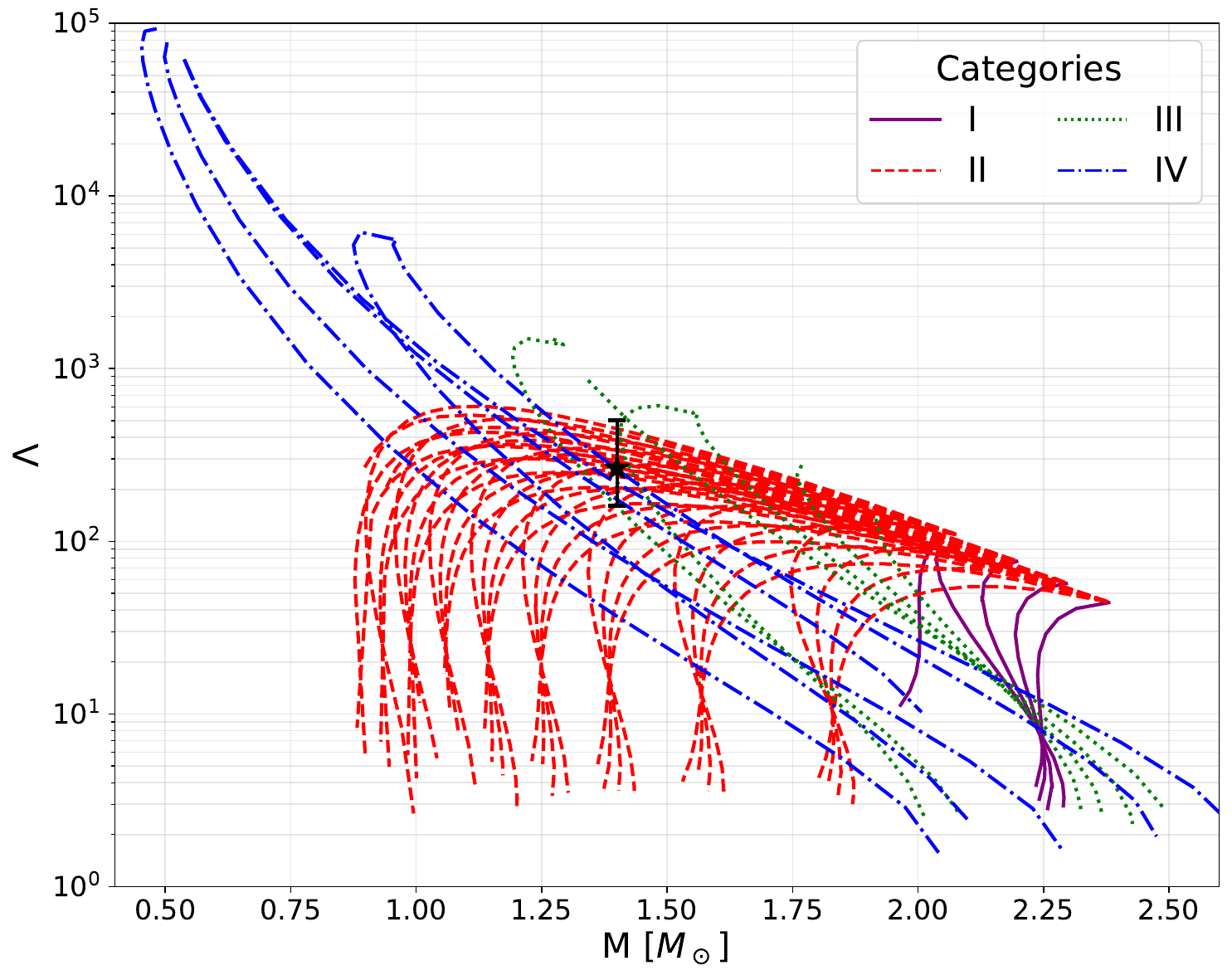}
    \caption{Dimensionless tidal deformability, $\Lambda$, for the SSHS as a function of the gravitational mass, $M$ for all the categories considered in this work. The colors used are the same as previous figures. We present in black the constraint for the $\Lambda$ of a $1.4\,M_\odot$ NS, i.e., $\Lambda(1.4M_\odot)\equiv \Lambda_{1.4}$, obtained in Ref.~\cite{Huang:2025mrd}.}
    \label{fig:masa-lambda}
\end{figure}
\begin{figure*}
    \centering
    \includegraphics[width=0.9\linewidth]{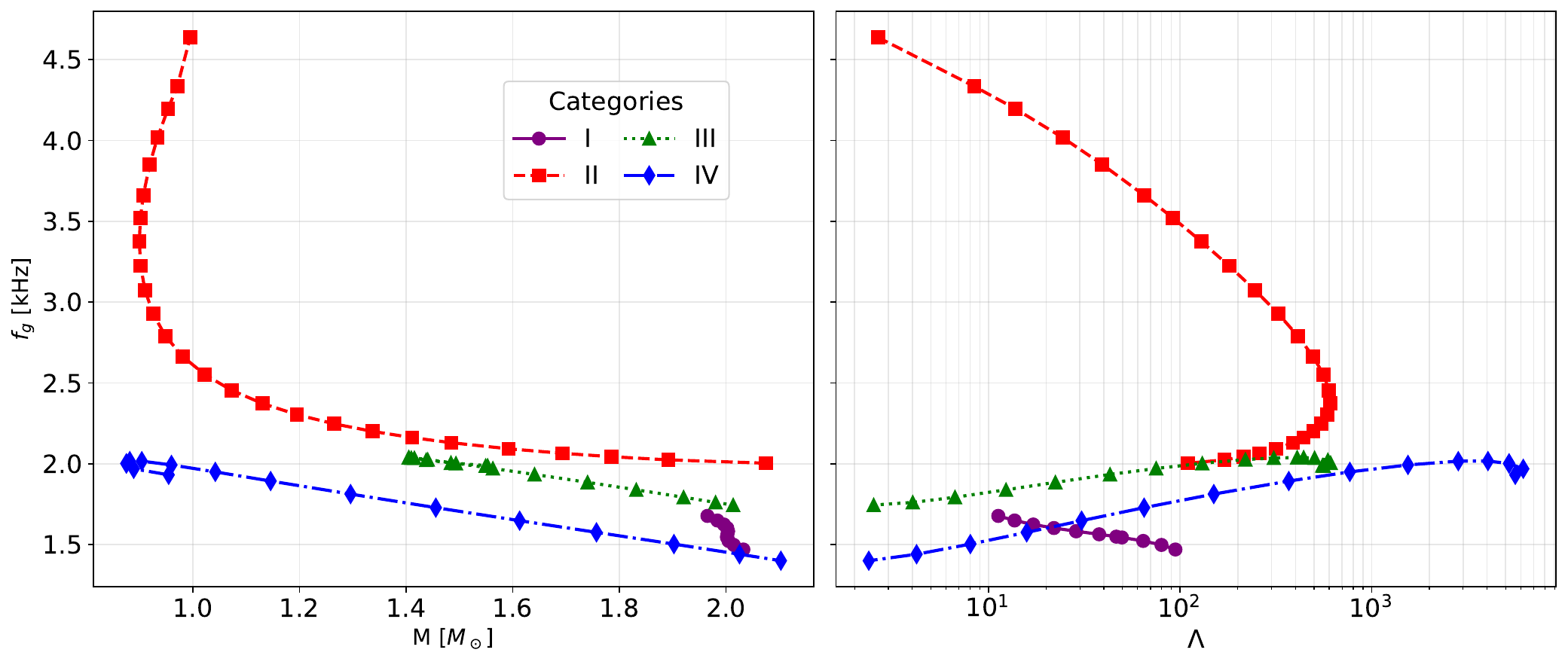}
    \caption{Comparison of the non-radial $g$-mode frequencies, $f_g$, as functions of the stellar mass, $M$, (left panel) and dimensionless tidal deformabilities, $\Lambda$, (right panel), displayed for one representative EoS per hybrid star category (I–IV). Each one of them is depicted with a distinct line style and marker for clarity. 
    }
    \label{fig:frec-masa-lambda}
\end{figure*}
In addition, it is relevant to analyze the behavior of the tidal deformability as a function of the stellar mass. In Fig. \ref{fig:masa-lambda}, we display $\Lambda(M)$ for all four categories of HSs considered in this work. The same color code as in previous figures is used to distinguish between categories, while the black star (with the error bars) corresponds to the constraint on the tidal deformability of a $1.4~M_{\odot}$ NS, i.e. $\Lambda(1.4M_\odot)\equiv \Lambda_{1.4}$, reported in Ref.~\cite{Huang:2025mrd}. We note that not all equations of state within each category satisfy this constraint. For Category I, we note that no SSHS configurations were obtained around $1.4~M_{\odot}$, $\Lambda_{1.4}$, and therefore the $\Lambda_{1.4}$ constraint cannot be applied to this case. Category II, on the other hand, produces the largest spread, with several models exceeding the allowed range of $\Lambda_{1.4}$. Categories III and IV produce values closer to the constraint, but some EoSs yield values of $\Lambda_{1.4}$ that fall outside the constraint. This figure therefore highlights how the specific choices of transition pressure and energy-density discontinuity affect non-trivially the tidal properties of SSHS.

\begin{figure}
    \centering
    \includegraphics[width=0.9\linewidth]{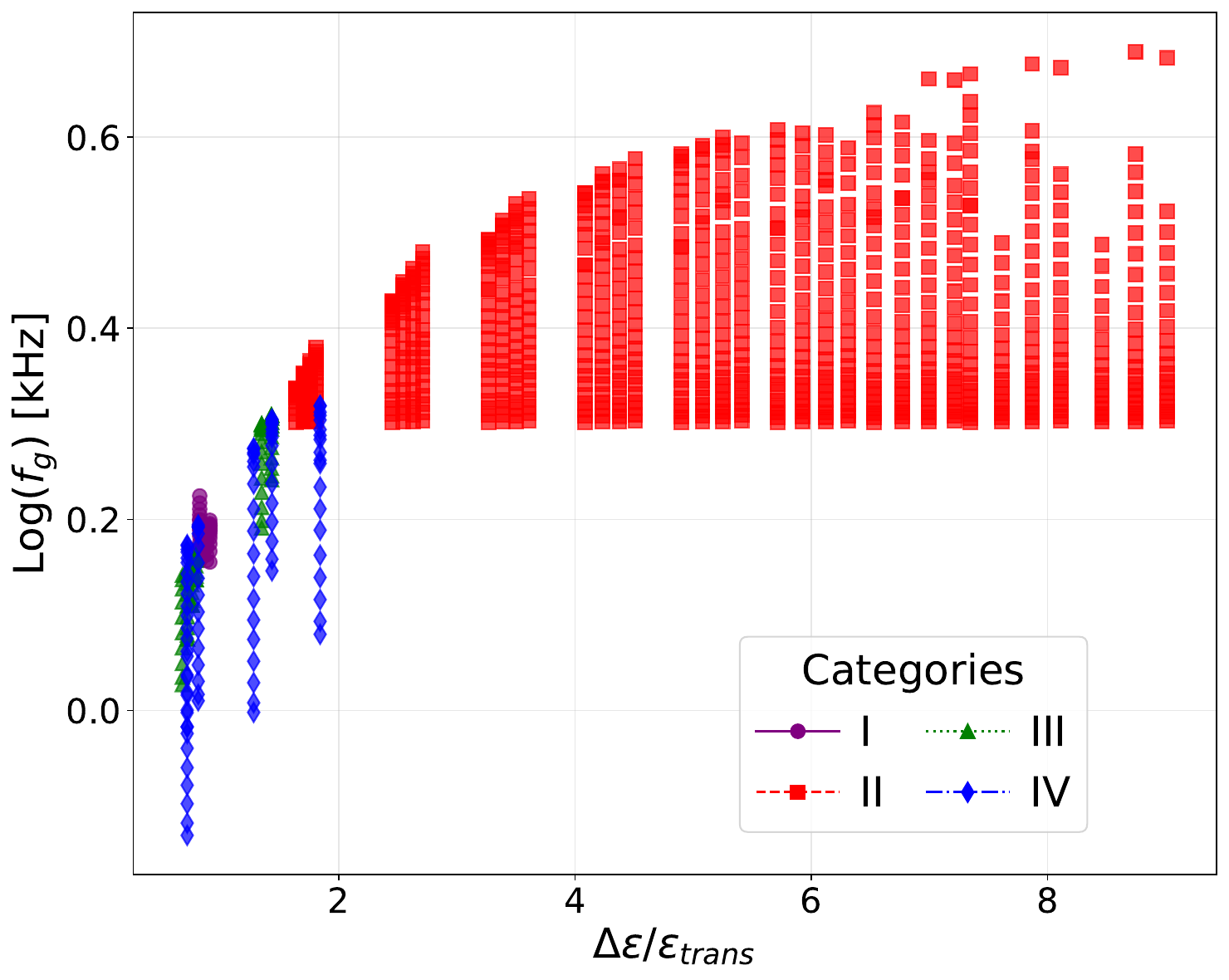}
    \caption{Relationship between the decimal logarithm of the $g$-mode frequencies (in kHz) and the relative energy density variation, $\Delta \epsilon / \epsilon_{\rm trans}$, for all  hybrid star EoSs in each category (I–IV) analyzed in the present work. Points represent individual stellar configurations from our analysis, color-coded by {EoS} category.
    }
    \label{fig:eps-frec}
\end{figure}

Now, after presenting details of our results for each category, we compare the $g$-mode frequencies, $f_g$, across the different categories in order to probe their qualitative and quantitative differences as a function of the gravitational masses and tidal deformabilities. In Fig.~\ref{fig:frec-masa-lambda}, we present these different behaviors for one representative EoS of each category (I–IV). It is revealed in the left panel that Categories I, III, and IV exhibit relatively flat trends, with frequencies between approximately 1.4 and 2.1 kHz across the mass range. In contrast, Category II shows a {quite} different qualitative and quantitative behavior, with $f_g$ increasing rapidly for decreasing values of $M$ and reaching values as high as $\sim 5$ kHz. We have seen that this novel trend happens because Category II stars have larger $\Delta\epsilon$ and $P_t$ that produce such non-trivial behavior compared to the other categories and past works exploring this non-radial mode. This is also manifested in the right panel of Fig.~\ref{fig:frec-masa-lambda} for $f_g$ against the tidal deformabilities where the non-monotonic behavior can be seen. 
In this sense, one can say that Category II hybrid stars are the most extreme scenario with a few low-mass configurations reaching high values of $f_g \sim 4-5$ kHz, i.e. more challenging to be detectable but still having a unique behavior compared to standard NSs and other hybrid-star categories, thus making them easily distinguishable in potential upcoming measurements if the mass of the pulsating object could be estimated from independent observations. If not, it would not be easy to discern its nature as other modes of NSs lie in the same range of frequencies. Besides, the fact that these discontinuity $g$-mode frequencies, $f_g$, are surprisingly large allows us to reliably discriminate them from the standard compositional $g$-modes with frequencies below 1 kHz \cite{Zhao:2022toc,Constantinou:2023ged}. On the other hand, we stress that apart from this small set of large-$f_g$ stars, most of the others lie in the range $f_g \sim 2-4$ kHz, in the range in which the third generation GW detectors present their best sensitivity, making them potentially detectable if sufficiently excited by tidal forces \cite{Evans:2021gyd}. Estimating the overlap integrals (see, Ref. \cite{counsell:2025imi} and references therein) of such modes might be key to address the issue of detectability more carefully. This quest is left for a future work. 

\section{Universal relations for discontinuity \texorpdfstring{$g$}{g}-modes} \label{sec:res2}

We have seen in the previous section that $g$-mode frequencies for the four categories are highly sensitive to the sharp energy density jump (or to the energy density jump normalized by the energy density at the transition) at the quark-hadron interface present in the EoSs for hybrid stars. This motivates us to probe whether a set of universal relations exists between the frequencies $f_g$ and $\Delta \epsilon / \epsilon_{\text{trans}}$ across the wide set of allowed EoSs of Subsec. \ref{subsec:eos}, thus allowing us to classify them by category.

Interestingly, we found that, contrary to what has been proposed in Refs.~\cite{ranea:2018omo,Rodriguez:2021hsw}, a relation (using a fitting function like Eq. \eqref{eq:fit1}) between the decimal logarithm of $f_g$ and $\Delta \epsilon / \epsilon_{\text{trans}}$ does not hold for all EoSs within each category (I–IV). We display these results in Fig.~\ref{fig:eps-frec} where a trend relating larger values of $\Delta \epsilon/\epsilon_{\text{trans}}$ to larger values of $f_g$ appears to exist. Each point corresponds to an individual hybrid star configuration, color-coded differently for each EoS category. 
\begin{equation}
{\rm Log_{10}} (f_g)= A + B~{\rm Log}_{10}(\Delta \epsilon / \epsilon_{\rm trans}),
\label{eq:fit1}
\end{equation}

\begin{figure}
    \centering
    \includegraphics[width=0.9\linewidth]{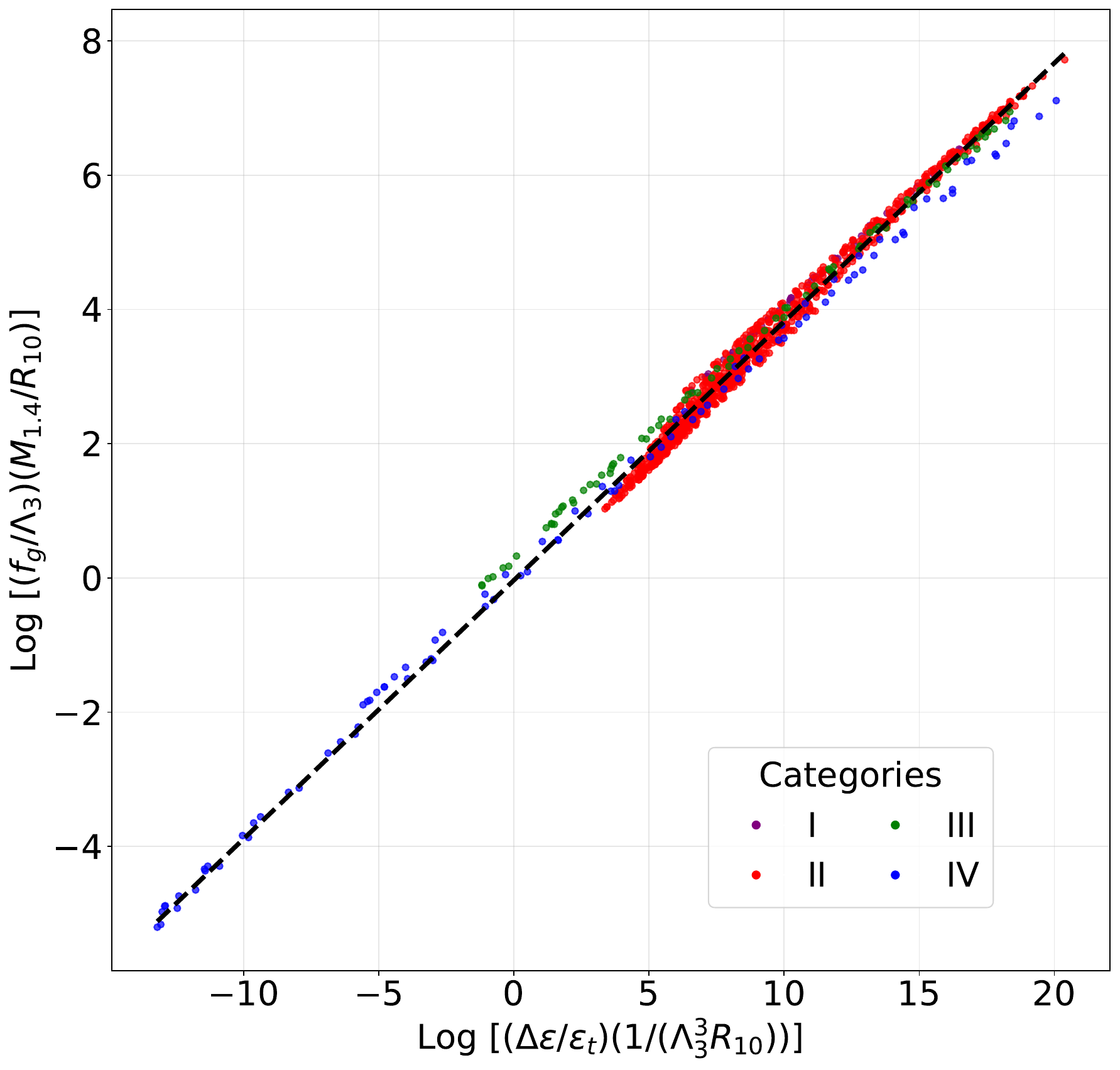}
    \caption{Proposed novel universal relation for hybrid stars of Eq.~\eqref{eq:fit2}. Points represent individual stellar configurations from our analysis, color-coded by EoS category. The dashed black line shows a linear fit to these data, demonstrating a tight correlation across hybrid star categories. 
    }
    \label{fig:ur-g}
\end{figure}

Due to the lack of universality of the previously proposed relations, especially when long branches of SSHS are taken into account, it would be interesting to investigate the dependence of the $g$-mode frequencies on the features of the phase transition as well as their structural properties. In order to do this, we examine the potential existence of a unique universal relation, i.e. the same for the four categories of hybrid stars, that combines $f_g$, tidal deformability, $\Lambda$ and stellar compactness of a canonical {NS} to the normalized energy density jump. We discovered that such a relation exists, and it is given by:
\begin{eqnarray} \label{eq:fit2}
{\rm Log}_{10}\left[(f_g / \Lambda_3)(M_{1.4} / R_{10})\right]=\\ \nonumber & \hspace{-3.5cm} C+D~{\rm Log}_{10}\left[(\Delta \epsilon / \epsilon_{\rm trans}) (1/( \Lambda_3^3R_{10}))\right],
\end{eqnarray}
where $M_{1.4} \equiv M / 1.4 M_{\odot}$, $R_{10} \equiv R / 10$ km, and \mbox{$\Lambda_3 \equiv \Lambda / 10^3$}. This relation can be thought of as a generalization of the one presented in Eq.~\eqref{eq:fit1} including effects related to macroscopic magnitudes of the compact objects that play a key role, particularly the long branches of SSHS that appear when EoSs of Category II are taken into account. In Fig.~\ref{fig:ur-g} we display this novel relation, where each point corresponds to a stellar configuration from our dataset, being color-coded by EoS category. The dashed black line represents the best linear fit across all categories, with slope $D = 0.385 \pm 0.001$ and intercept $C = -0.036 \pm 0.011$. The high Pearson correlation coefficient ($0.9961$) 
indicates a robust universal trend linking $g$-mode frequencies to phase transition parameters present in the EoS and global stellar properties. 
Most points lie within a few percent of the fit, reinforcing the accuracy and predictive power of the proposed universal relation. We close this section by highlighting that our proposed universal relations in this work are the first attempt in the literature to go beyond the usual relations between $f_g M$ and $\Lambda$ \cite{Kuan:2022bhu}{. These conventional relations} do not always produce reliable results approaching the universal curve, often leading to relatively large relative errors compared to those established relations for the $f$-mode \cite{Kuan:2022bhu}, a mode that is currently under intense research. 

\section{Summary and Outlook} \label{sec:conc}

In this work, we explored the quantitative and qualitative behavior of the non-radial $g$-mode frequencies (directly associated with the first-order phase transition) for the four categories of  hybrid stars. In order to do this, we used the CSS parametrization of the hybrid EoS to match a hadronic matter phase onto a quark matter phase. The hadronic phase is characterized by a combination of the BPS (at very low densities) and SLy EoS at low baryon densities between $\sim$ 0.5 $n_0$ and 1.1 $n_0$, i.e. the region of validity of the CET results. From this density onward, we used three generalized piecewise polytropes (GPP) until the transition point, which, depending on the category of  hybrid stars, is located at different values of $ \lbrace P_t,~ \Delta\epsilon \rbrace$ and $c^{2}_s\in [0.5,1]$ of the quark phase. Equipped with these inputs, we solved the TOV equations to obtain the $M$-$R$ diagrams for each of the four categories of hybrid stars. Then, applying criteria of Ref.~\cite{Pereira:2018pte} for SSHS, we found that all our stellar families (hadronic and hybrid) branches are dynamically stable against radial oscillations.

These results were employed to solve the relativistic quasi-normal non-radial oscillation equations to determine the $g$-mode frequencies, $f_g$, in the relativistic Cowling approximation. After careful numerical calculations, we found the explicit dependencies of $f_g$ on the stellar mass, $M$ and tidal deformability, $\Lambda$. Apart from expected quantitative differences for results between categories, strongly related to the triad $ \lbrace P_t,~ \Delta\epsilon,~c^2_s \in [0.5,1] \rbrace$, we discovered a global explanation for their qualitative peculiarities. 

We also explored the exciting possibility of building universal relations for: a) each category of HSs and b) a unique formula encompassing the four categories. For a), following the reasoning of Ref.~\cite{Rodriguez:2021hsw}, we found four relations that, despite having small correlation coefficients, connect ${\rm Log_{10}} (f_g)$ to $\Delta\epsilon/\epsilon_{\rm trans}$. Interestingly, they display notoriously distinct slopes, departing from the 1/2 result obtained with Newtonian gravity.  {The low-correlation factors found using previously proposed universal relations of this type reinforce one of our main conclusions: previous proposals of universal relations break down when slow stable hybrid stars are taken into account,  {something well known for other non-radial modes \cite{Ranea:2022bou,Ranea:2023auq}}.} For b), based on our calculations for the $g$-mode frequencies of Sec. \ref{sec:res1}, we proposed a logarithmic relation between the $f_g$, the compactness of a canonical NS, the normalized energy density jump and the tidal deformability. This linear formula found unified results for all the categories with very low values of relative error. This gives us confidence to use it to extract information about the phase transition in terms of usually measured values of compactness and tidal deformabilities of NSs. Such relations are, therefore, highly relevant for inferring non-radial frequencies through gravitational wave data, as has already been done for the non-radial $f$-mode frequencies \cite{Pratten:2019sed}.

Moreover, if a simultaneous detection of a $g$-mode and the fundamental $wI$-mode occurs, the universal relations from Ref.~\cite{Ranea:2023cmr} could be used to constrain the star's mass and radius (within a few percent) and its dimensionless tidal deformability (within a factor of 2). These estimations would then serve to determine $\Delta \epsilon / \epsilon_t$ via the universal relation established in this work.

 {A comment is in order on the asteroseismological applicability of our proposed universal relation. In order to be able to estimate $\Delta \epsilon / \epsilon_t$, not only the frequency of the $g$-mode needs to be estimated but also values of several macroscopic quantities of the pulsating compact object  {should be known with good accuracy}. This is of course not an easy quest, but when the Cosmic Explorer and Einstein telescope-network becomes functional, a rate of pre-merger (\textit{i.e.} GW emission during the very early stages of the inspiral phase) detections of NS-NS mergers of the order of tens per year, with good spatial localization is expected (see, for example, Ref. \cite{nitz:2021pml}). This information would help better determine the sky localization creating imrproved alerts for telescopes to be able to detect the electromagnetic counterpart of the merger. Moreover, according to several works, the number of binary-NS (BNS) mergers detected by the third-generation gravitational-wave detectors is expected to be, in an optimistic scenario, $\sim 10^5$ BNS mergers per year (see, for example, Ref. \cite{singh:2022ecb,Abac:2025tso,bisero:2026mmo} references therein). This, together with potential electromagnetic counterparts detections, would allow restricting macroscopic quantities such as the mass, the radius, and the dimensionless tidal deformability of NSs.  {Furthermore,} new facilities are expected to produce, depending on the nature of the EoS, radius estimations with an accuracy of about 30 m for neutron stars in the 1 to $\sim 2\,M_\odot$ mass range \cite{huxford:2024acn}. At the same time, the chirp mass and the binary’s mass-weighted tidal deformability would be greatly constrained. Therefore, the masses of merging objects  {will be potentially} measured to better than a few percent  {of accuracy} \cite{puecher:2023mte}.  {Since these} quantities appear in our proposed universal relations for the discontinuity $g$-modes,  {it is expected that {after these improvements, the proposed relation might} be tested and perhaps further constrained in the near future}. This might render our proposed universal relation a useful tool in the next decade of multimessenger astronomy. Our goal is therefore not to claim immediate applicability, but rather to establish a theoretical tool that can be  {consistently applicable} once sufficiently accurate post-merger measurements become available.}

Of course, another interesting question to be addressed is related to the detectability (directly or indirectly) of our proposed $g$-mode classes (and associated universal relations) from current and upcoming gravitational wave data. 

Firstly, it is interesting to discuss the energetics needed in order to estimate the potential detectability of these $g$-mode frequencies of hybrid stars in future gravitational-wave detectors (like the Advanced LIGO-Virgo and Einstein telescopes). The frequencies of the $g$-modes associated with compact configurations of Categories I, III, and IV are comparable to those studied in Refs. \cite{tonetto:2020dgm,Rodriguez:2021hsw}. For this reason, we expect a similar energy threshold. The situation changes when objects of Category II are taken into account. For this family, the $g$-modes span a larger range of frequencies where, in addition, gravitational-wave detectors do not present their best sensitivities. Despite the need of a more detailed analysis, we expect that for these objects, a larger amount of energy has to be channeled into gravitational waves so that they become detectable.

 {Closely related to this subject, the results presented in Ref. \cite{counsell:2025imi} are interesting. The authors conclude that the shift in the orbital phase due to resonant tidal excitation of $g-$modes related to sharp phase transitions might be detectable even with LIGO A+. The general trend that can be seen from Fig. 2 of Ref. \cite{counsell:2025imi} is that the phase shift increases both with the frequency of the mode and with $\Delta \epsilon / \epsilon_t$. Having present that the sensitivity curve at 5 kHz is only 20 $\%$ worse than for 2 kHz \cite{capote:2025ald}, this effect could be detectable for SSHSs, particularly for Category II stars. The potential detectability would dramatically increase when Cosmic Explorer and the Einstein Telescope begin to operate, as expected.}

Interestingly, a recent complementary research presented in Ref.~\cite{2025arXiv250416911P} pointed out the difficulty of reliably distinguishing $f$-mode frequencies from the $g$-mode in hybrid stars. This challenge arises from their simultaneous coupling with the dynamical tidal field, producing an overlap between their numerical values for stellar masses from 1.4 to 2 $M_{\odot}$. However, they only considered a fixed value of $P_t$ while making slight variations of the transition jump, $\Delta \epsilon$ maximally reaching values around 1. However, we believe that a similar study must be carried out considering strong first-order phase transitions for which we have seen that the $g$-mode frequencies behave quite differently (quantitatively and qualitatively) for the four categories of  hybrid stars, something that lies beyond the scope of the present work and is left for future works. 
\begin{acknowledgments}
 The authors thank Luciano Rezzolla for useful discussions on his proposed bound on compactness. Thanks also to the anonymous referees for the constructive comments and criticisms about the detectability and usefulness of our findings as well as his/her suggestions to improve the quality and presentation of our results. M.C.R is a doctoral fellow at CONICET (Argentina). M.C.R and I.F.R-S. acknowledge UNLP and CONICET (Argentina) for financial support under grants G187 and PIP 0169.  J.C.J. is supported by Conselho Nacional de Desenvolvimento Científico e Tecnológico (CNPq) with Grant No. 151390/2024-0. The work of I.F.R-S. was also partially supported by PIBAA 0724 from CONICET.
\end{acknowledgments}
\appendix
\section{Explicit solutions for the $g$-modes of the $V$ and $W$ eigenfunctions} \label{App:A}

\begin{figure}[h!]
    \centering
    \includegraphics[width=0.99\linewidth]{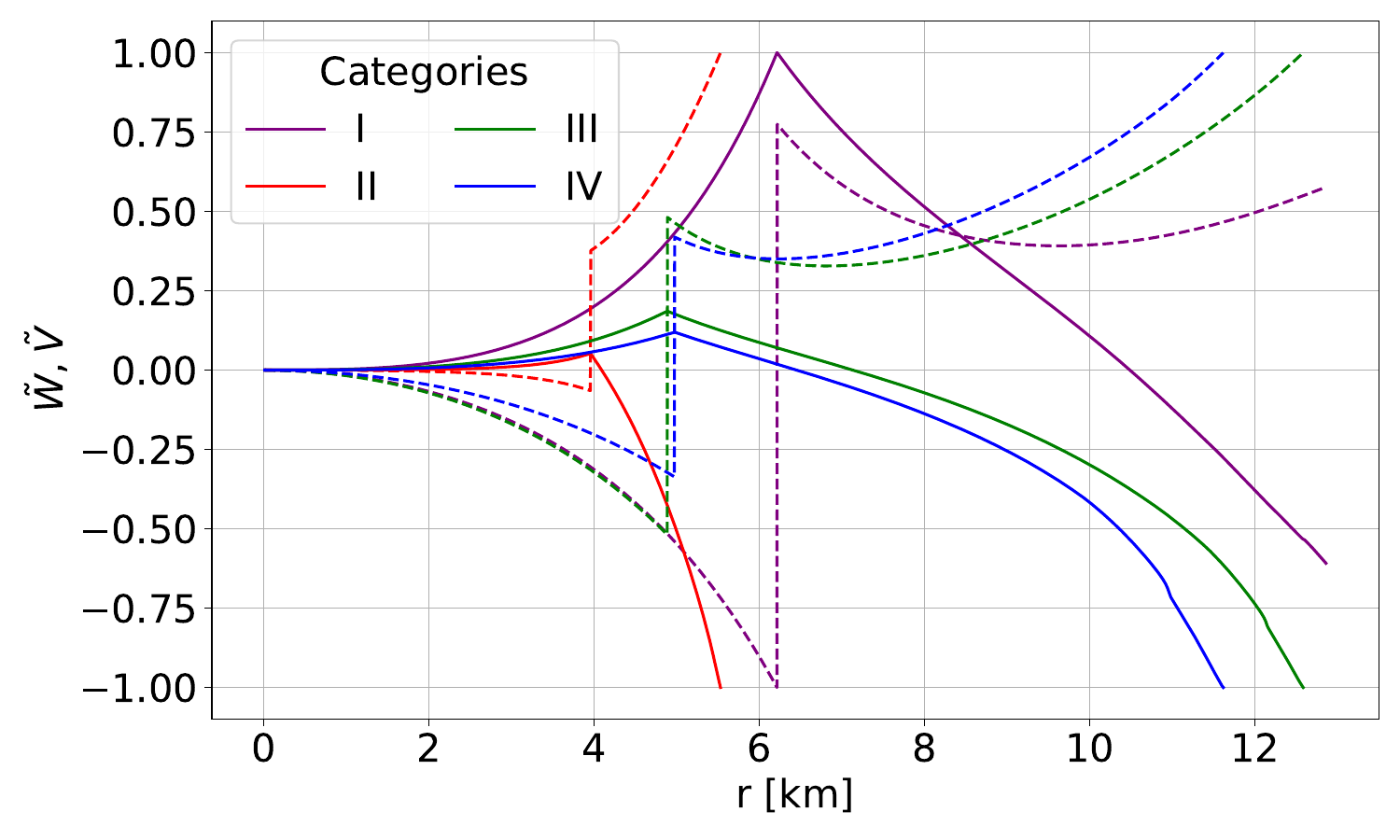}
    \caption{Radial profiles of the normalized (by their respective maximum amplitudes) eigenfunctions $\tilde{W}(r)$ (solid lines) and $\tilde{V}(r)$ (dashed lines) for the $g$-mode, computed for stellar configurations representative of Categories I-IV. The same color is used for both eigenfunctions within each category. The values of the CSS used to calculate these particular models are presented in Table \ref{tableexample}.}
    \label{fig:autofunciones}
\end{figure}

Radial eigenfunctions $V$ and $W$ for the $g$-mode for representative cases of each of the four categories are shown in Fig.~\ref{fig:autofunciones}. Each curve is calculated for the terminal SSHS configuration obtained using a hybrid EoS constructed with the CSS parameters presented in Table \ref{tableexample}. Every curve is normalized by its respective maximum amplitude{, i.e. $\tilde{V}$ and $\tilde{W}$}, and consistent colors are used to represent both eigenfunctions within each category. The discontinuity observed in the {$\tilde{V}$} eigenfunction occurs at the hadron-quark phase transition in the core of the star. Similarly, the eigenfunction {$\tilde{W}$} exhibits an abrupt change in its radial derivative at the location of the phase transition. Related to this point, we also analyzed the behavior of the $f$-mode for these four hybrid EoS, i.e. the functions $\tilde{W}$ and $\tilde{V}$, as well as their frequencies. We found that the $f$-mode frequencies are pushed to larger values due to the presence of our discontinuous $g$-modes. This allows us to safely distinguish $g$ from $f$ modes due to their distinct scale of frequencies. Interestingly, this agrees with results of Refs. \cite{tonetto:2020dgm,Mariani:2022omh,Ranea:2023auq}. Thus, our distinction between the $f$-mode and $g$-modes based upon just on the number of nodes for the $\tilde{W}$ and $\tilde{V}$ functions becomes consistent.

\begin{table}
\begin{tabular}{| c | c | c | c | }
\hline 
Category & $\Delta \epsilon$ & $P_t$       & $c^2_s$   \\ \hline \hline 
I        & $274$       &  $70$ & $1$  \\
II       & $2446$      &  $70$ & $1$   \\  
III      & $405$       &  $40$  & $1$    \\ 
IV       & $346$       &  $20$  & $1$      \\ \hline   
\end{tabular}
\caption{{Parmeters used to calculate the example eigenfunctions show in Fig. \ref{fig:autofunciones}. The units used for the different quantities are the same as in Table \ref{tableEoSs}.}}\label{tableexample}
\end{table}

\section{Tidal deformability in neutron stars with a first-order phase transition} \label{App:B}
The calculation of the dimensionless tidal deformability, $\Lambda$, follows the formalism outlined below. For quadrupolar perturbations, $\Lambda$ is related to the Love number, $k_2$, which can be obtained by solving, together with the TOV equations, an additional differential equation:
\begin{eqnarray} \label{eqz}
r\frac{{\rm d}\zeta (r)}{{\rm d}r} &+& \zeta (r)^2 + \zeta (r)\frac{1+4\pi r^2 \left[P(r) + \epsilon (r)\right]}{1-\frac{2m(r)}{r}} \nonumber \\
&-& \frac{6 - 4\pi r^2 \left[5\epsilon (r) + 9P(r) + \frac{\epsilon (r) + P(r)}{{\rm d}P/{\rm d}\epsilon}\right]}{1-\frac{2m(r)}{r}} \nonumber \\
&-&\left(r\frac{{\rm d}\nu (r)}{{\rm d}r}\right)^2 = 0 \, , 
\end{eqnarray}
with the boundary condition $\zeta (0) = 2$. Furthermore, for {EoSs} with abrupt discontinuities at radii $r=r_{\rm t}$, the additional matching condition
\begin{equation}
\zeta ({r_ {\rm t}}^+) - \zeta ({r_{\rm t}}^-) = \frac{4\pi {r_{\rm t}}^3 \left[\epsilon ({r_{\rm t}}^+) - \epsilon ({r_{\rm t}}^-)\right]}{m({r_{\rm t}}) + 4\pi (r_{\rm t})^3P(r_{\rm t})} ,
\end{equation}
must be imposed \citep{Takatsy2020, Han:2019tdw}. The superscripts $+$ and $-$ represent the values at the transition radius on either side of the interface. After solving Equation~\eqref{eqz}, we compute the Love number, $k_2$, using the following expression,
\begin{eqnarray}
k_2 &=& {\frac{8}{5}\beta^5(1-2\beta)^2[2+2\beta(\zeta-1)-\zeta]}\nonumber \\
&\times& \Big[2\beta[6-3\zeta+3\beta(5\zeta-8)] \nonumber \\
&+& 4\beta^3[13-11\zeta+\beta(3\zeta-2)+2\beta^2(\zeta+1)] \\ 
&+& 3(1-2\beta)^2[2-\zeta+2\beta(\zeta-1)]\ln(1-2\beta)\Big]^{-1} \ , \nonumber
\end{eqnarray}
where $\zeta \equiv \zeta(R)$. The quantity $\beta = M/R$ denotes compactness. The dimensionless tidal deformability is then obtained using the following expression,
\begin{equation} \label{dimtidal}
\Lambda = \frac{2}{3}k_2\left(\frac{R}{M}\right)^5 \ .
\end{equation}

\bibliography{references}

\end{document}